\PassOptionsToPackage{table}{xcolor}

\documentclass[acmtog]{acmart}

\renewcommand\footnotetextcopyrightpermission[1]{} 
\acmSubmissionID{616}
\setcitestyle{nosort,square} 

\AtBeginDocument{%
  \providecommand\BibTeX{{%
    \normalfont B\kern-0.5em{\scshape i\kern-0.25em b}\kern-0.8em\TeX}}}

\usepackage{color}
\usepackage{xcolor}
\usepackage{xspace}
\usepackage{wrapfig}
\usepackage{placeins}
\usepackage{afterpage}
\usepackage{dblfloatfix} 
\usepackage{enumitem}

\hypersetup{
    colorlinks=true,
    linkcolor=blue,
    urlcolor=red,
    linktoc=all,
    citecolor=black
}

\makeatletter
\DeclareRobustCommand\onedot{\futurelet\@let@token\@onedot}
\def\@onedot{\ifx\@let@token.\else.\null\fi\xspace}

\def\eg{\emph{e.g}\onedot} 
\def\ie{\emph{i.e}\onedot}

\newif\ifshowedits

\newcommand{\addeditor}[3]{%
  \definecolor{#1color}{rgb}{#3}
  \expandafter\newcommand\csname #1\endcsname[1]{%
  \ifshowedits
    {\color{#1color} ##1}%
  \else
    {##1}%
  \fi
  }%
  \expandafter\newcommand\csname #1rmk\endcsname[1]{%
  \ifshowedits
    {\color{#1color} {\bf [#2: ##1]}}
  \fi
  }%
  \expandafter\newcommand\csname #1rpl\endcsname[2]{%
  \ifshowedits
    {{\color{#1color} ##1} \sout{##2}}
  \else
    {##1}
  \fi
  }%
}

\definecolor{darkgreen}{RGB}{0,110,0}
\definecolor{darkred}{RGB}{170,0,0}

\addeditor{amir}{AB}{0.0, 0.0, 0.0}
\addeditor{tg}{TG}{0.0, 0.0, 0.0}
\addeditor{vk}{VK}{0.0, 0.0, 0.0}
\addeditor{noam}{NA}{0.0, 0.0, 0.0}
\addeditor{amit}{AB}{0.0, 0.0, 0.0}

\addeditor{todo}{td}{1.0, 0.0 0.0}
\showeditstrue

\newcommand{\ourmethod}{MagicClay}

\begin{document}

\title{\ourmethod{}: Sculpting Meshes With Generative Neural Fields}

\author{Amir Barda}
\email{amirbarda@mail.tau.ac.il}
\affiliation{%
  \institution{Tel Aviv University}
  \country{Israel}
}

\author{Vladimir G. Kim}
\affiliation{%
  \institution{Adobe Research}
  \country{USA}}

  \author{Noam Aigerman}
\affiliation{%
 \institution{Université de Montréal}
 \country{Canada}}

\author{Amit H. Bermano}
\affiliation{%
  \institution{Tel Aviv University}
  \country{Israel}}

\author{Thibault Groueix}
\affiliation{%
  \institution{Adobe Research}
  \country{USA}
}

\begin{abstract}


The recent developments in neural fields have brought phenomenal capabilities to the field of shape generation, but they lack crucial properties, such as incremental control --- a fundamental requirement for artistic work. Triangular meshes, on the other hand, are the representation of choice for most geometry-related tasks, offering efficiency and intuitive control, but do not lend themselves to neural optimization. 
To support downstream tasks, previous art typically proposes a two-step approach, where first, a shape is generated using neural fields, and then a mesh is extracted for further processing. Instead, in this paper, we introduce a hybrid approach that maintains both a mesh and a Signed Distance Field (SDF) representations consistently. Using this representation, we introduce \ourmethod{} --- a tool for sculpting regions of a mesh according to textual prompts while keeping other regions untouched. \amir{Our method is designed to be compatible with existing mesh sculpting workflows. The user sculpts the desired shape using the existing brushes and our pipeline then evolves the geometry and triangulation of the selected mesh part according to the given textual prompt. This process operates on the original mesh while preserving its meta-data}
Our framework carefully and efficiently balances consistency between the representations and regularizations in every step of the shape optimization. Relying on the mesh representation, we show how to render the SDF at higher resolutions and faster. In addition, we employ recent work in differentiable mesh reconstruction to adaptively allocate triangles in the mesh where required, as indicated by the SDF.
Using an implemented prototype, we demonstrate superior generated geometry compared to the state-of-the-art and novel consistent control, allowing sequential prompt-based edits to the same mesh for the first time. 
\tg{We will release the code upon acceptance.}
\end{abstract}

%
%



\begin{teaserfigure}
\centering
  \includegraphics[width=0.95\textwidth]{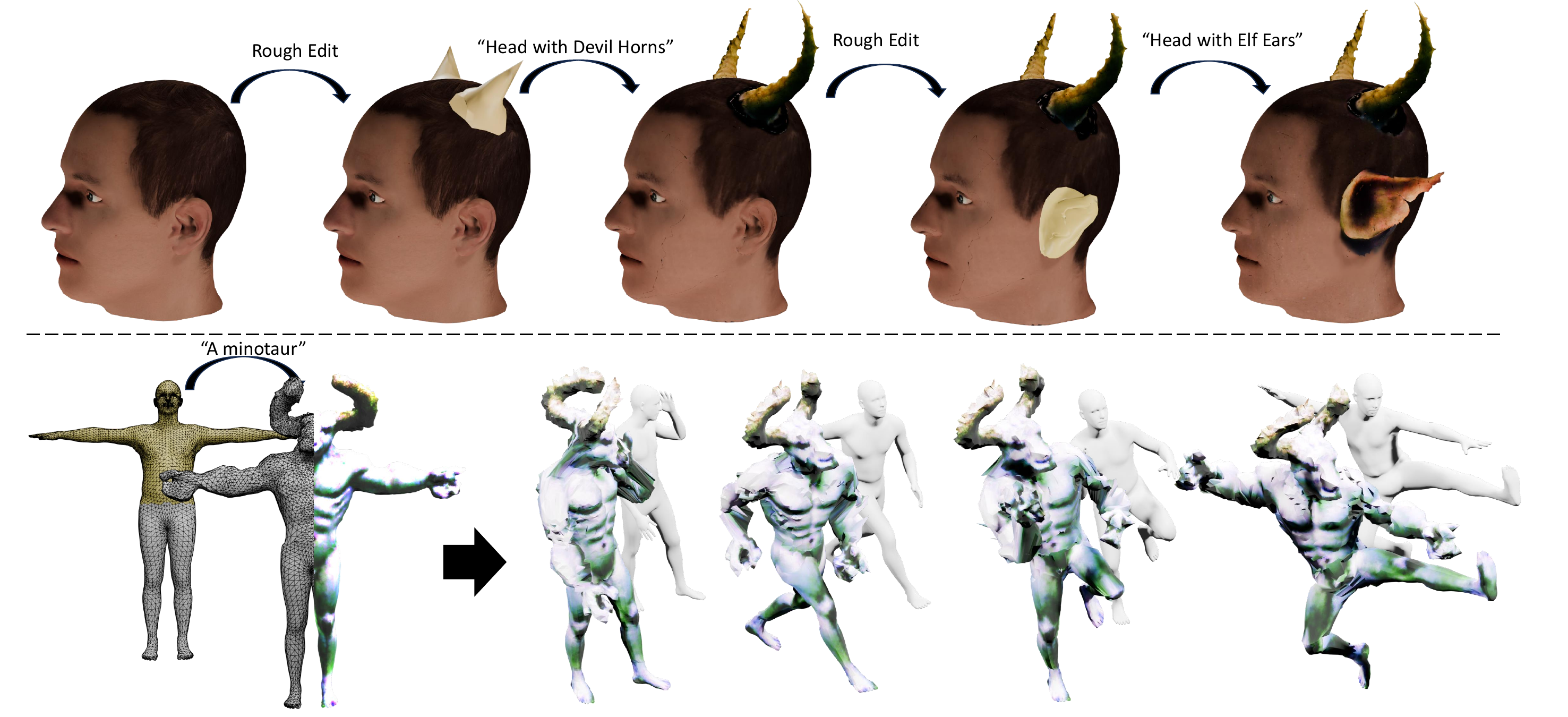}
  \caption{\textmd{
  \textit{Top:} the user selects a region of an input mesh, coarsely sketches the intended edit, inputs a text-prompt, and \ourmethod{} grows the region automatically to match the prompt, while the rest of the shape and existing textures remain unchanged, and the mesh remains topologically valid (no non-manifold edges or vertices). Importantly, due to working directly on the input mesh, these edits can be made sequentially. \textit{Bottom:} \ourmethod{} preserves attributes from the input mesh such as texture, rig, and tessellation, \eg allowing to transfer an animation from source to edited shape. 
  }}
  \label{fig:teaser}
\end{teaserfigure}


\maketitle

\section{Introduction}

The field of 3D shape generation has always been heavily dependent on the representations it uses for the shapes.
Recent neural field-based representations (i.e., NeRFs~\cite{mildenhall2021nerf} or SDFs~\cite{deepsdf, chen2018implicit_decoder}), have shown remarkable progress to the task~\cite{poole2022dreamfusion, wang2023prolificdreamer} in a very short time.
These representations are robust to noisy losses and are naturally well-suited for neural frameworks, yielding impressive results and avoiding local minima. On the other hand, these representations are expensive to evaluate (limited by volumetric rendering resolutions) and lack acutely in geometric control, \vk{such as allowing users to localize their edits or leverage surface-based (e.g., smoothness) priors}. For instance, when iterating on a design of a mesh-based 3D model, as artists alter geometry, textures, or topology, they expect any additional updates to retain previously assigned attributes. This level of control is currently not feasible with the neural field-based representations.

In contrast, triangular meshes provide such control and are indeed the current dominant representation for most 3D applications in the industry, as they are computationally inexpensive, consistent, and intuitive.   On the other hand, while the adaptive, non-uniform nature of meshes is one of their greatest advantages, it is also the reason they are not widely used in current generative frameworks. The sparse gradients induced by meshes tend to limit the ability of optimizations to achieve large deformations in a stable manner.

For this reason, many works~\cite{lin2023magic3d} turn to a two-step process, where first implicit functions are used for coarse generation, and then are converted to meshes in a second step for the purpose of finer details or downstream editability. However, as we demonstrate, two-stage pipelines are prone to local minima and cannot be extended to the task of editing an existing mesh with pre-existing UVs and textures, for example. 

In this paper, we present \ourmethod{} - a shape editing framework based on a hybrid implicit-explicit representation. \ourmethod{} optimizes an \amir{input mesh (possibly textured)} and \amir{its} SDF jointly at every step of the generation process and leverages consistency and representation-specific priors, benefiting from the best of both worlds. This approach leads to higher quality generation output and also enables a higher level of control, allowing the users to sequentially sculpt local generative details \amir{in regions marked as editable, guided by the user-provided prompt}. 
Sculpting is a common approach used in 3D modeling software~\cite{blender,zbrush, SubstanceModeler}. While sculpting currently requires a lot of time and expertise, our novel tool offers unprecedented control by allowing artists to select a region on a mesh to be modified, provide a textual prompt, and hallucinate an updated region (Figure~\ref{fig:teaser}). 

The key technical challenge of the hybrid 
approach is keeping the two representations synced efficiently. To achieve this, we differentiably render both representations from various angles and require consistency in RGB renders, opacity, and normal maps. Furthermore, we rely on the in-sync-mesh representation to render the SDF at higher resolutions and faster; instead of the hundreds of samples per ray required in most previous techniques~\cite{poole2022dreamfusion}, we localize the SDF sampling around the mesh surface and use as little as three samples. 
Critically, optimizing a mesh consistently and stably is an additional challenge. In terms of resolution, a coarse mesh would not be expressive enough for novel details, and a fine mesh is expensive and unstable. Hence, an adaptive tessellation is required, that changes along with the shape where needed.  
We rely on recent developments in differentiable mesh reconstruction~\cite{barda2023roar} to achieve dynamic mesh topology updates, including face splitting, edge collapse, and edge flips. To texture the mesh despite changes in its topology, we contribute a new strategy based on triangle supersampling. Importantly, using this layer, we can maintain mesh properties (e.g, vertex groups for animation) throughout the optimization, and through local mesh topology changes.
 
As we demonstrate, our hybrid approach allows localized and sequential mesh editing operations, allowing the user to preserve existing mesh triangulation and information in some regions, while allowing radical and semantic changes in other regions. In addition, we show overall higher generated geometric quality as compared to using an implicit representation alone, thanks to the priors the two representations impose on each other. 
\vk{By combining the merits of implicit and explicit representations to enable a novel generative sculpting tool, we bring the neural shape generation pipeline closer to the artistic workflows, allowing for incremental editing steps and providing the artist with precision and control over the end result.}


\tg{
To summarize, our main contributions are :
\begin{itemize}[left=0pt, topsep=0pt]
    \item A new hybrid representation that brings the benefits of SDFs and meshes together : SDFs are more robust to noisy gradients (see Figure~\ref{fig:noiserobustness}) and meshes allow for surface-based losses and localization of edits. Both representations are kept consistent via multi-view consistency losses and adaptive remeshing. In Figure~\ref{fig:gen_comparison} and Table~\ref{tab:quantitative}, we demonstrate in the task of unconditional text-to-3D  that \ourmethod{} produces higher-quality meshes than other representations using similar generative techniques.
    \item An application of the new representation to localized mesh editing, which brings the prowess of generative techniques to mesh sculpting, while critically preserving the part of the geometry that is \textit{not} selected for editing, including its texture, tessellation, and rigging parameters. We show in Figure~\ref{fig:comparison} that our method produces higher-quality and more localized edits than competing methods.
\end{itemize}
}

\label{sec:intro}
\begin{figure*}

\includegraphics[width=0.9\textwidth]
{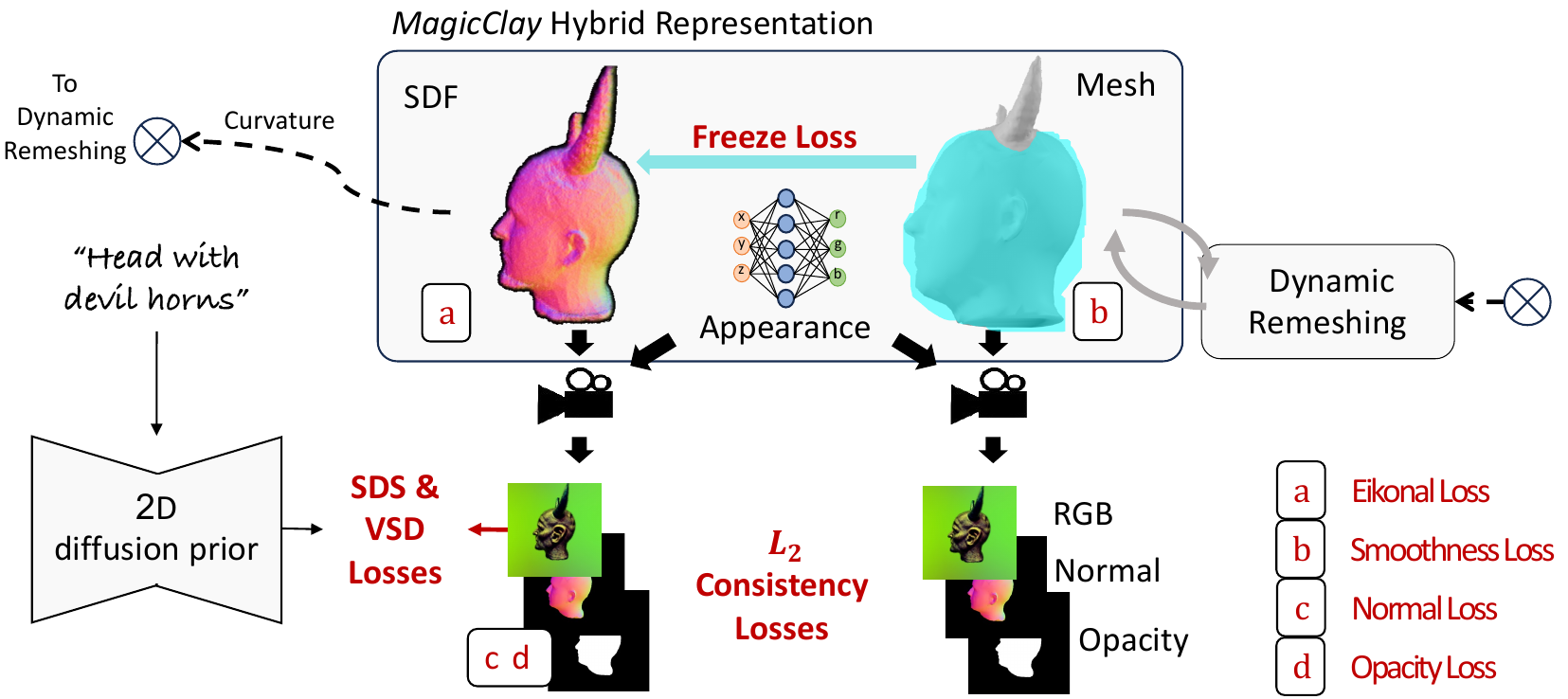}

\caption{\bf{Overview of the hybrid optimization.} \textmd{We jointly optimize a mesh, an SDF and a shared appearance MLP according to an input prompt. We can either optimize the full geometry, or only a user-selected portion of the mesh for an iterative 3D modeling workflow. We can also preserve existing textures on non-selected part of the mesh, or have the diffusion model generate textures for the full mesh. We start by differentiably rendering both representations, and enforcing their consistency. As they are kept in sync, we use the mesh to efficiently sample volumetric rays to render hi-res maps from the SDF in a memory-efficient manner. Applying SDS-type losses on these Hi-res renderings allows for capturing finer details.
We sync the mesh and the SDF via multi-view consistency constraints on the RGB pixels, the image opacity, and the surface normals.
The mesh local topology is updated according to the SDF using ROAR~\cite{barda2023roar}, splitting triangles where geometry is created and collapsing edges where needed. Additionally, we leverage representation-specific losses to regularize the optimization: an Eikonal loss on the SDF and a smoothness loss on the mesh. }}

\label{fig:ROAR_overview}
\end{figure*}

\section{Related Work}
\paragraph{{\bf Unconditional 3D generative models.}}
In their seminal work, DreamFusion,~\cite{poole2022dreamfusion} show that Text-to-Image diffusion models can be used to provide gradients to optimize a  Neural Radiance Field (NeRF) via Score Distillation Sampling (SDS). Magic3D \cite{lin2023magic3d} achieves better quality by using a two-stage approach: the first stage is similar to DreamFusion, and they note that the quality of the generated object is limited by the high cost of performing volumetric rendering for high-resolution images. The second stage uses a differentiable mesh representation to refine the generated object further, as differentiably rendering meshes in high resolution is significantly cheaper in both time and memory. Magic123 \cite{qian2023magic123} further improves upon Magic3D by using both 3D and 2D diffusion priors. ProlificDreamer \cite{wang2023prolificdreamer} proposes an improvement over SDS, the VSD loss,  to drive 3D generation from 2d diffusion priors. Fantasia3D \cite{chen2023fantasia3d} and TextMesh \cite{tsalicoglou2023textmesh} decouple the appearance from the geometry by replacing the NeRF with an SDF, and optimizing a color network separately. TextDeformer \cite{gao2023textdeformer} uses CLIP as prior together with a novel gradient smoothing technique based on mesh Jacobians to deform meshes according to a text prompt.

The choice of using two stages in Magic3D~\cite{lin2023magic3d} highlights the tradeoffs involved in choosing the right representation for 3D generative models. While implicit functions are well suited for coarse generation because they allow topology updates, meshes can be rasterized very efficiently at a high resolution  to get fine details in a second step. However, two-stage pipelines are prone to local minima and crucially, cannot be extended to edit an existing mesh with pre-computed UVs. In contrast, we jointly optimize a hybrid SDF and mesh representation, that can be initialized from an existing mesh and maintain all of its properties during optimization, benefiting from the best of both worlds in a \textit{1-step pipeline} : topology updates from the SDF part and fine details from the mesh part.

\tg{Finally, Instant3D~\cite{instant3d2023} generates a 3D shape in a single forward pass without requiring any costly optimization. Though faster and higher-quality, Instant3D does not allow sculpting an existing mesh like \ourmethod{}.
}
\paragraph{{\bf Generative local editing.}}

\tg{A recent line of work deals with locally editing a 3D scene. Unlike 2D images, selecting a region in 3D to be edited is not straightforward for implicit functions. Most of these works solely rely on soft attention between the text and the renderings to semantically understand the area to be edited, and they vary with their choice of 3D representation. 
Vox-E~\cite{sella2023voxe} and DreamEditor~\cite{zhuang2023dreameditor} encode an SDF respectively with a voxel grid and an MLP. Instruct-NerftoNerf~\cite{instructnerf2023}, LatentNerf ~\cite{metzer2022latent}, SKED~\cite{mikaeili2023sked}, MVEdit~\cite{mvedit2024} use a NeRF. 
In these methods, the user selects the region to edit via the prompt, except for SKED~\cite{mikaeili2023sked}, which additionally utilizes guiding sketches from different views. 
\amir{
FocalDreamer \cite{li2023focaldreamertextdriven3dediting} and Progressive3D \cite{cheng2024progressive3dprogressivelylocalediting} use an implicit representation with SDS and additionally allow selection of 3D region(s) with losses encouraging localized changes. The output mesh is reconstructed from the edited SDF using DMTeT, thus losing all mesh properties in the non-editable regions.}
In contrast, our method exposes a mesh to the user, on which it is trivial to select a region and guarantees that the unselected part will be unaltered.
} \newline

{\bf Hybrid Representations.}  unlike for 2D images, there is no ubiquitous representation in 3D, , and several representations exist and have been combined for diverse 3D tasks, suggesting that there is no one-fit-for-all solution. In this work, we introduce a hybrid representation specialized for generative modeling and focus the related work on hybrids most relevant to this paper.
~\cite{OmidHybrid2020} uses a coupling of implicit and explicit surface representations for generative 3D modeling, kept in sync by 3D losses.  NerfMeshing \cite{rakotosaona2023nerfmeshing} proposes an improved meshing pipeline for NeRFs.  
Finally, DmTeT~\cite{shen2021dmtet} proposes deep marching Tetrahedra  as a hybrid representation for high-resolution 3D Shape synthesis, notably used in the concurrent work Magic3D~\cite{lin2023magic3d}. Our method uses both a set of regularization losses, as well as a dynamic projection layer based on ROAR~\cite{barda2023roar} to keep the SDF and mesh part in sync. \newline

{\bf Traditional approaches for sculpting meshes}
Many commercial tools employ the digital sculpting metaphor for 3D modeling, such as Zbrush~\cite{zbrush}, Mudbox~\cite{mudbox}, or SubstanceModeler~\cite{SubstanceModeler}. Motivated by these workflows, geometry processing research has focused on improving interactive techniques such as mesh deformation~\cite{skinning_survey_2014}, mesh blending for cut-and-paste~\cite{Biermann_2002_cut_and_paste}, local parameterization for adding surface details~\cite{Schmidt_expmaps2006}, symmetry-guided autocompletion~\cite{Peng_autocomplete_sculpting}, and version control for collaborative editing~\cite{Salvati2015_meshHisto}. Despite these advances, 3D modeling remains only accessible to experts. As an alternative, example-based approaches propose to democratize 3D modeling tools by using existing geometry from a database of stock 3D models to assemble new shapes from parts~\cite{Funkhouser2016}. Subsequent methods have built statistical models over part assemblies~\cite{Kalogerakis:2012:ShapeSynthesis}, and allow semantic control for deformations~\cite{yumer_2015_semantic}. Despite their accessibility, these tools are often restricted in their domain, and rely on heavy annotation of 3D assets, and thus have received limited use by professional modelers. In this paper, we utilize pre-trained 2D generative data priors to enable semantic controls for local and iterative modeling workflow without the need for pre-annotated 3D data.

\section{Method}
\label{sec:method}

Given a mesh, a user-highlighted surface region, and a text prompt that describes the desired target, \ourmethod{} optimizes the shape of the selected region so that the resulting mesh matches the target. 
To drive the shape optimization, we follow current literature and use the Score Distillation Sampling (SDS) technique \cite{poole2022dreamfusion} with differentiable rendering to leverage on text-conditioned 2D diffusion and guide the shape optimization. This approach, however, does not perform well when operated on meshes. Meshes are driven by sparse and irregular samples (vertices), and their connectivity mandates a stable and smooth deformation, avoiding self-intersections and flip-overs. For this reason, we employ a neural Signed Distance Field (SDF) to drive the mesh shape optimization and topology updates. We thus propose a hybrid representation that captures both a Signed Distance Field (SDF) and the surface, gaining from the advantages of both worlds. While the SDF allows guiding the shape toward larger-scale complex changes, the mesh allows localized control of the user-highlighted surface region. 

In this section, we provide details on the hybrid SDF/Mesh representations (Sec.~\ref{subsec:representation}), how it can be efficiently optimized with SDS guidance (Sec.~\ref{subsec:guidance}), how to effectively use surface and volumetric priors (Sec.~\ref{subsec:priors}), and how to update the mesh topology during optimization (Sec.~\ref{subsec:topology}). Figure~\ref{fig:ROAR_overview} overviews the full pipeline.


\subsection{Hybrid Representation}
\label{subsec:representation}

Our hybrid representation consists of a surface (a mesh), a volume (an SDF), and a shared appearance network encoding RGB colors for an input 3D coordinate. Both the surface and the volumetric representations can be differentiably rendered, leveraging the shared appearance network to output images with color, normals, and opacity channels.  We now detail these three elements.

\paragraph{Surface Representation}
We represent the surface of the shape as a 2-manifold triangular mesh. Mesh topology, or sampling resolution, is locally adapted according to the SDF (see Sec.~\ref{subsec:topology} for details).  
We encode colors for the editable areas of the mesh using an auxiliary appearance network, derived from the SDF itself (see below). We found this approach simpler and more natural than traditional mesh coloring techniques; 
Using per-vertex colors is sensitive to triangulation, and would require a large number of vertices to match the resolution of the SDF. Using a texture image requires a complex UV parameterization, usually done a priori on a fixed shape. In addition, our surface is continuously optimized and undergoes topological changes, re-tessellation, and large-scale deformations, which makes it computationally infeasible to apply traditional UV parameterization techniques during this optimization.

Instead, our hybrid approach offers a simpler approach to shape coloring. To apply the colors from the appearance network to the mesh, we propose to adaptively subdivide each face of the base mesh according to the triangle area. 

Since we only use these subdivided triangles to represent colors, they do not have to form a connected mesh, unlike traditional subdivision techniques. Thus, we employ the MeshColors scheme that was initially proposed for UV-less texturing~\cite{meshcolors2010} and has an efficient GPU implementation.

In the inset, we illustrate the example subdivision; note how sub-triangles on two adjacent faces do not share the vertices along the 
\setlength{\columnsep}{10pt}%
\begin{wrapfigure}{r}{0.2\textwidth}
  \centering
  \includegraphics[width=0.9\linewidth]{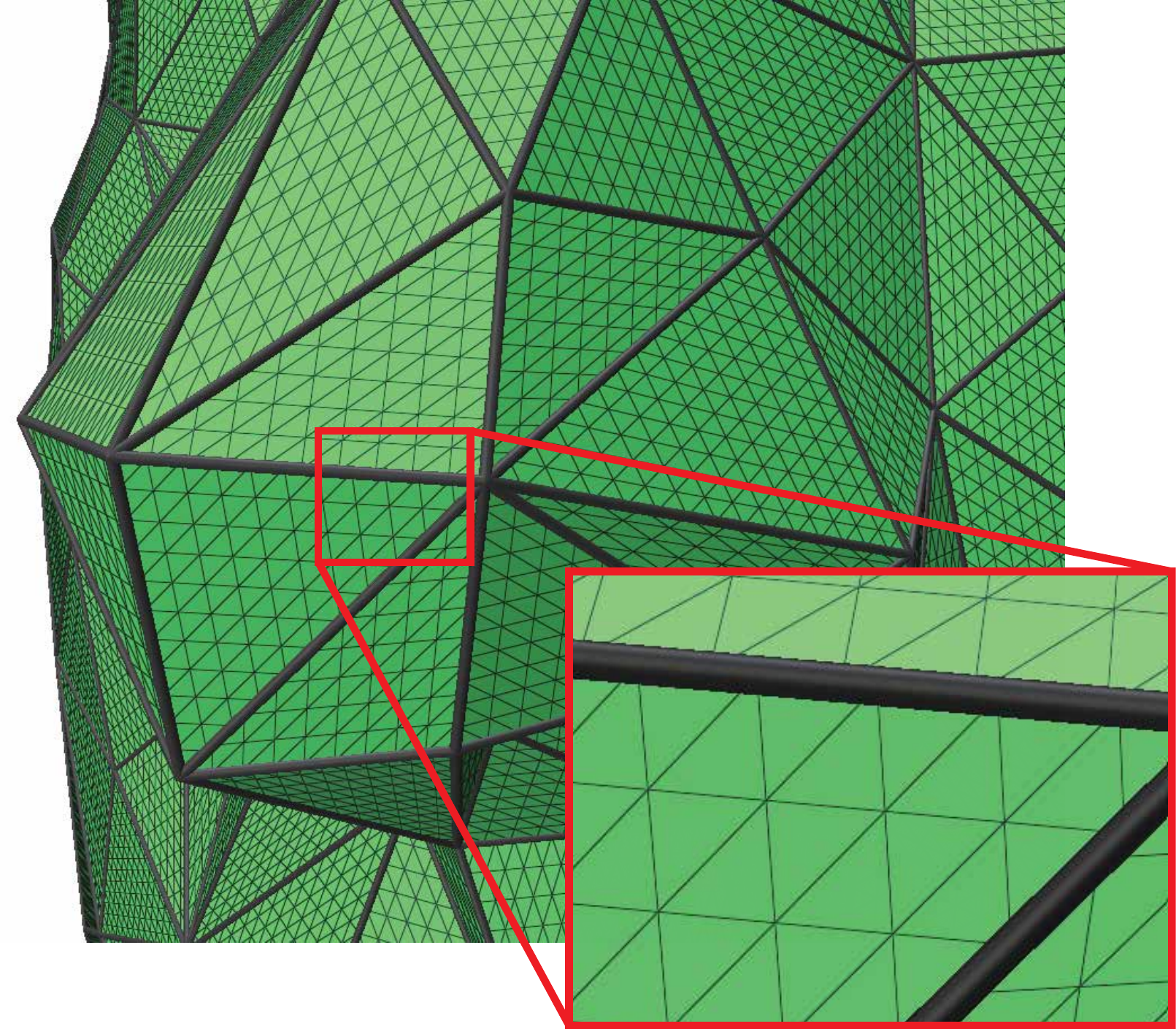}
\end{wrapfigure} 
edge. \tg{During rendering, we assign a color to each sub-triangle by interpolating the colors at its three vertices. These are obtained by querying the appearance network at the super-sampled vertices location.}. \amir {We find that this approach strikes a balance between a compact low-poly representation for flat parts of the generated surface, while still allowing for high-frequency textures.} 
%
%
%


\paragraph{Signed Distance Functions} 
Our volumetric shape representation is chosen off-the-shelf, and conceptually serves as a regularization guiding the mesh \vk{changes} using existing state-of-the-art text2shape tools. We use a continuous scalar field that can be sampled anywhere in $\mathbf{R}^3$, returning a signed shortest distance to the surface (negative on the inside, positive on the outside). We encode the SDF using a multiresolution hash encoding of features defined over a grid which are then mapped to distance value by a small MLP, following instant-NGP~\cite{instantngp}.
~As in the mesh case, the shared appearance network is sampled to obtain colors during rendering.


\paragraph{Appearance Network} 
The shared appearance network encodes colors implicitly as a map over $\mathbf{R}^3$.
It shares the same hash grid as the SDF, but has a smaller MLP head, with a single hidden layer that take hash grid features as input and outputs RGB values.  

\subsection{Hybrid Shape Guidance}
\label{subsec:guidance}
Our shape optimization is based on Score-Distillation Sampling (SDS) to distill gradients from a text prompt. 
The primary motivation to maintain an SDF representation in addition to the mesh is because SDFs are more robust to noisy guidance, which is an inherent 
property of the multi-view SDS approach (see Figure~\ref{fig:noiserobustness}). We thus choose to inject the text guidance only to the auxiliary SDF representation and propagate the changes to the mesh via the consistency losses (Sec.~\ref{subsec:priors}) and the topology updates (Sec. ~\ref{subsec:topology}).


To apply the text guidance and the consistency losses, we need to render both representation differentiably. We use Nvdiffrast~\cite{nvdiffrast} to render meshes and VolSDF~\cite{volsdf} for volumetric rendering of our SDF. Clearly, as mesh rasterization is much cheaper than volumetric rendering, the process is bottlenecked by the resolution at which we can render the SDF, both in terms of speed and memory.
Our hybrid representation uniquely enables a strategy to render SDF faster and cheaper, at a higher resolution of 512x512.

This is achieved thanks to the consistency between the mesh and SDF representations throughout the optimization.

\setlength{\columnsep}{10pt}%
\begin{wrapfigure}{r}{0.2\textwidth}
  \centering
  \vspace{-1em}\includegraphics[width=\linewidth]{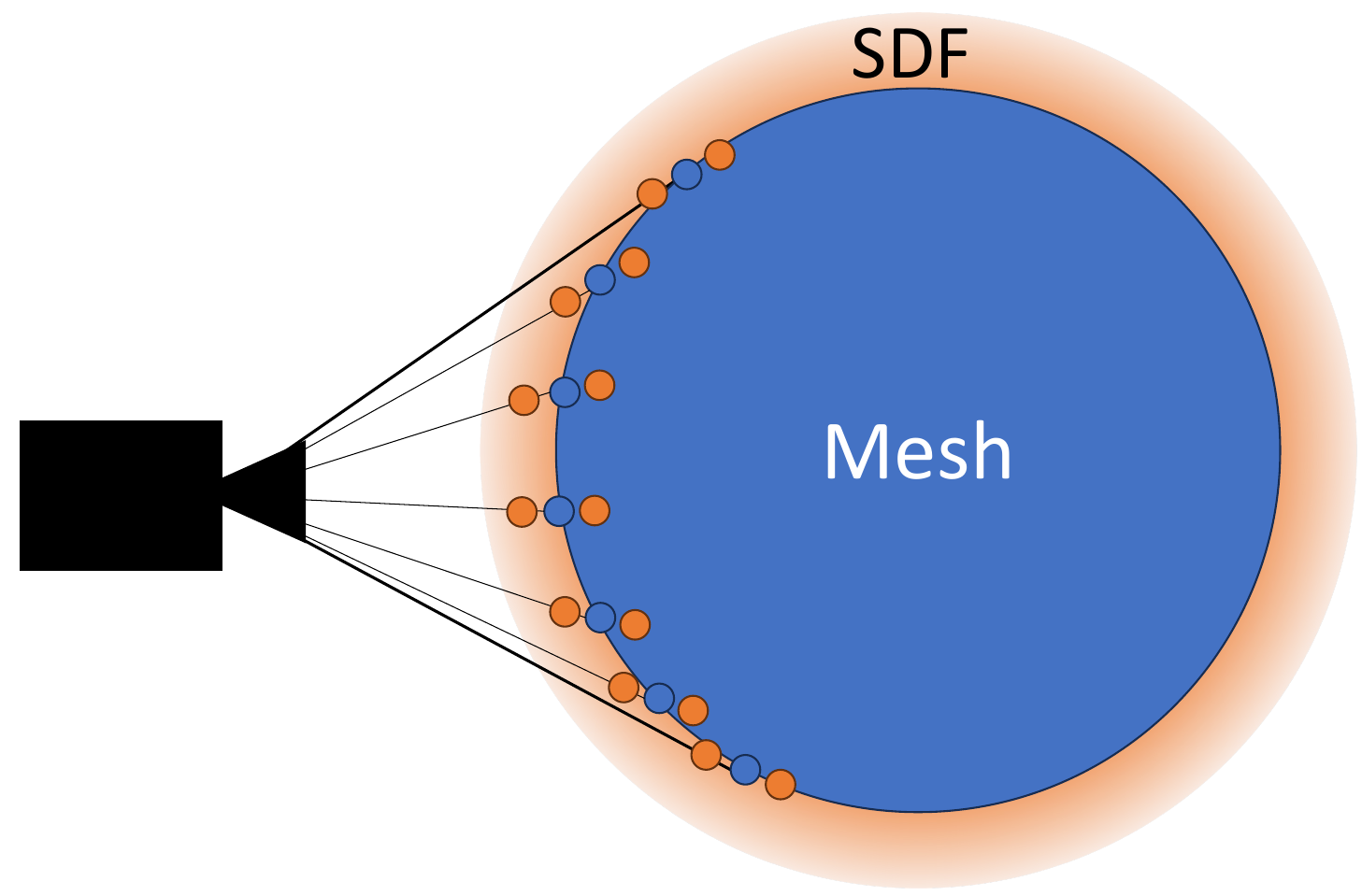}
  \vspace{-5mm}
\end{wrapfigure}
We can significantly reduce the typical 512 samples per ray necessary for rendering the SDF by using the intersection of the ray with the mesh representation (efficiently calculated by the differentiable mesh renderer). Using the intersection as the center of a small spread of samples (we use just 3 per ray), allows for high-resolution renders of the SDF (i.e., 512x512 and larger), which are otherwise memory prohibitive.
The idea of leveraging the surface to reduce the number of network queries per ray emerged in concurrent works, namely Adaptive Shell~\cite{adaptiveshells2023} and HybridNerf~\cite{turki2023hybridnerf}, which shows its generality and success in other settings than ours. 

Using this strategy, we render the SDF in 512x512 and apply the VSD loss of those high-res renderings. We also apply regular SDS on lower-res 128x128 renderings by regular VolSDF as we find that this improves the results slightly. 



\subsection{Representation Priors}
\label{subsec:priors}

We apply representation-specific regularizations and consistency losses that keep both representations in sync.

\paragraph{Consistency Loss.} The SDF and the mesh are consistent if their images are in 1 to 1 correspondences from any camera angle. We thus supervise the L2 difference between their RGB renderings and normal and opacity maps. If the renderings are made at different resolutions, we downsize to the lower resolution before the L2 loss.

\paragraph{Enforcing Localization and Freeze Loss} To localize changes to the user-selected area, we first fix the mesh vertices in all non-selected regions during optimization by zeroing out gradients outside of user selection. While localization is harder to achieve for SDF, we add a sampling-based freeze loss, which favors regions around fixed vertices to remain unchanged:

\begin{equation}
    s(v_\text{sampled}) = 0
\end{equation}

where $v_\text{sampled}$ are vertices sampled uniformly over the faces which are not part of the optimization region selected by the user.

\paragraph{Laplacian (Smoothness) Loss}
While it is harder to regularize the surface of an implicit function to be smooth, the explicit representation of the mesh allows to easily define a smoothness term using the Laplacian of the mesh, \tg{inspired by~\cite{kanazawa2018learning}}, defined for each vertex:

\begin{equation}
    \delta(x_i) = x_i - \frac{\Sigma_{j=1}^{N}{x_j}}{N},
    \label{eq:laplacian}
\end{equation}

where $x_j$ are \tg{the N} neighbors of $x_i$. The Laplacian vector encodes local \tg{curvature} changes: a smooth mesh is defined by low Laplacian vectors. To encourage smoothness, we use a global loss:

\begin{equation}
    L_\text{smooth} = \Sigma_i{||\delta(x_i)||}.
    \label{eq:smooth_loss}
\end{equation}

\tg{We opt for the uniform Laplacian instead of the cotan Laplacian, as the latter is more sensitive to ill-conditioned triangles, which may appear during the dynamic mesh updates.}

\paragraph{SDF Eikonal Loss.}
To encourage the implicit function to learn a valid SDF representation, we use the Eikonal term as a loss\tg{~\cite{eikonal}}. The SDF $s$ is valid if and only if the loss in Eqn~\ref{eq:eikonal_loss} is $0$:

\begin{equation}
    L_\text{Eik} = \Sigma_x{(||\nabla s(x)||-1)^2}
    \label{eq:eikonal_loss}
\end{equation}


\paragraph{SDF opacity and normal Loss.}
Inspired by  TextMesh~\cite{tsalicoglou2023textmesh}, we also binarize the SDF opacity and apply a Binary Cross Entropy loss to encourage discrete 0 or 1 values. To penalize badly oriented normals of the implicit surface, we apply an L2 penalty to the dot product between the normal and the camera direction if it is negative.

\subsection{Updating the Mesh Topology}
\label{subsec:topology}
To maintain consistency between the mesh and SDF, it is necessary to perform local topology edits on the mesh in that increase or decrease mesh resolution where required. Continuous Remeshing~\cite{continous_remshing2022} pioneered such a local topology update approach by using the Adam optimizer state as a signal. While this approach works well in a multi-view reconstruction scenario, where the images are sharp and the camera parameters known, the noise involved in SDS makes the gradients, and by extension the Adam state, very noisy and unstable signal to trigger those operations. 
We turn to another work, ROAR~\cite{barda2023roar}, particularly well-tailored to our hydrid representation. Within this framework, we use the SDF as the signal to trigger mesh triangle splits. 

In a nutshell, for each triangle on the mesh, ROAR starts by supersampling the triangle into K sub-faces, and projects each sub-vertices on the 0-level set of the SDF $s$ using a projection operator: 
%
%
\begin{equation}
    P(x) =- s(x) \cdot \nabla s(x)
    \label{eq:projection}
\end{equation}
%

This projection results in a piece-wise linear surface that approximates the implicit surface closest to the initial triangle. The decision to split this triangle is based on the curvature score of this piece of projected surface. If the surface is very curved, then the triangle is split using $\sqrt{3}$-subdivision~\cite{subdivision}. 
Similarly, each edge is assigned a score based on the quadratic distance of its vertices to all the planes in the 1-ring of the edge, which intuitively represents how important the edge is to the geometry. If the score is low, then the edge can be collapsed with Qslim~\cite{qslim}.

We refer the interested reader to the ROAR paper~\cite{barda2023roar} for more details, but the important point to note is that ROAR offers a principled way to perform edge collapses and face splits in the sense that each iteration of ROAR strictly decreases an energy -  the difference between the highest face score and the lowest edge score. It thus exhibits a convergence behavior after enough iterations. We also note that manifoldness is guaranteed to be preserved throughout the iterations. 

\section{Experiments}
\label{sec:experiments}

\begin{table}
\label{tab:quantitative}
\centering
\scalebox{0.78}{
\begin{tabular}{lcc|c|cc}
 & &  & Appearance & \multicolumn{2}{c}{Geometry} \\
\cmidrule{4-4}
\cmidrule{5-6}
Average & & Implicit & Implicit & Implicit & Mesh no texture \\
CLIP Score $\uparrow$ & T2I & type & RGB  & Normal & Normal \\
 \hline
 ProlificDreamer & {SD2.1}&  NeRF& 22.1 & 21.2 
 & 20.1 
 \\
HIFA & {SD2.1} &  NeRF & 23.2 & 23.1 
& 22.0 
\\
MVDreams & {SD2.1}& NeRF & 25.9 & 25.4 
& 24.1
\\
\hline
Fantasia3D & {SD1.5}& SDF & - & 23.9 
&23.9 
\\
TextMesh & {DF}& SDF & 25.7 & 24.9 
& 23.6
\\
\rowcolor{yellow!20}\textbf{Ours}& DF & SDF   & \textbf{26.2} & \textbf{26.1
} 
&  \textbf{24.8
}
\\
  \hline
  {\color{teal}\scriptsize 1} {No high-res}& DF & SDF   & 25.8 & 25.2
  & 23.1
  \\
  {\color{teal}\scriptsize 2} {No low-res}& DF & SDF & 22.3
 & 21.9
  &  21.1
  \\
  \hline
\label{tab:quantitative}
\end{tabular}
}
\caption{\tg{{\bf Quantitative comparison on text-to-3D from scratch.} To validate our hybrid representation, we compare \ourmethod{} with five state-of the-art method for unconditional generation. We report the CLIP Score, \ie the cosine similarity (scaled by 100) between prompt embeddings and 25 normal renders of the NeRFs/SDFs and texture-less meshes, taken in a equidistant circular pattern around the generated object, on a benchmark of 20 prompts, listed in the supplementary. We apply marching cubes to extract the meshes for NeRF-based methods, and isosurface extraction for SDF-based methods, except ours, as our dual representation already includes a mesh. For each method, we report the Text-to-Image model (T2I) used as a backbone : DeepFloyd (DF) or Stable-Diffusion (SD). \amir{We Note that Fantasia3D can not be run with DF, as it does not support diffusion in pixel-space}. We validate that \ourmethod{} has the highest Clip score for texture-less meshes, which can be verified qualitatively in Figure~\ref{fig:gen_comparison}.
}}
\label{tab:quantitative}
\end{table}


We implement our pipeline in Threestudio~\cite{threestudio}, and use DeepFloyd  ~\cite{DeepFloyd} as the backbone diffusion model. All experiments were executed on a single A100-40GB GPU. We provide further details, hyper-parameters, and additional examples of the texture maps generated by our method and the animation transfer application in the supplementary material.

In the rest of this section, we compare our representation to prior work on text-conditioned 3D generation (Sec.~\ref{subsec:representation_comparison}), demonstrate its utility in a mesh sculpting application (Sec.~\ref{subsec:sculpting}), and compare to a text-driven mesh deformation baseline (Sec.~\ref{subsec:sculpting_comparison}).
We then provide a simple illustrative experiment to motivate the hybrid representation,
 and finally ablate our method (Sec.~\ref{subsec:ablations}).


\subsection{Comparison with Generative Methods}
\label{subsec:representation_comparison}

\begin{figure*}
    \centering
    \includegraphics[width=0.9\linewidth]{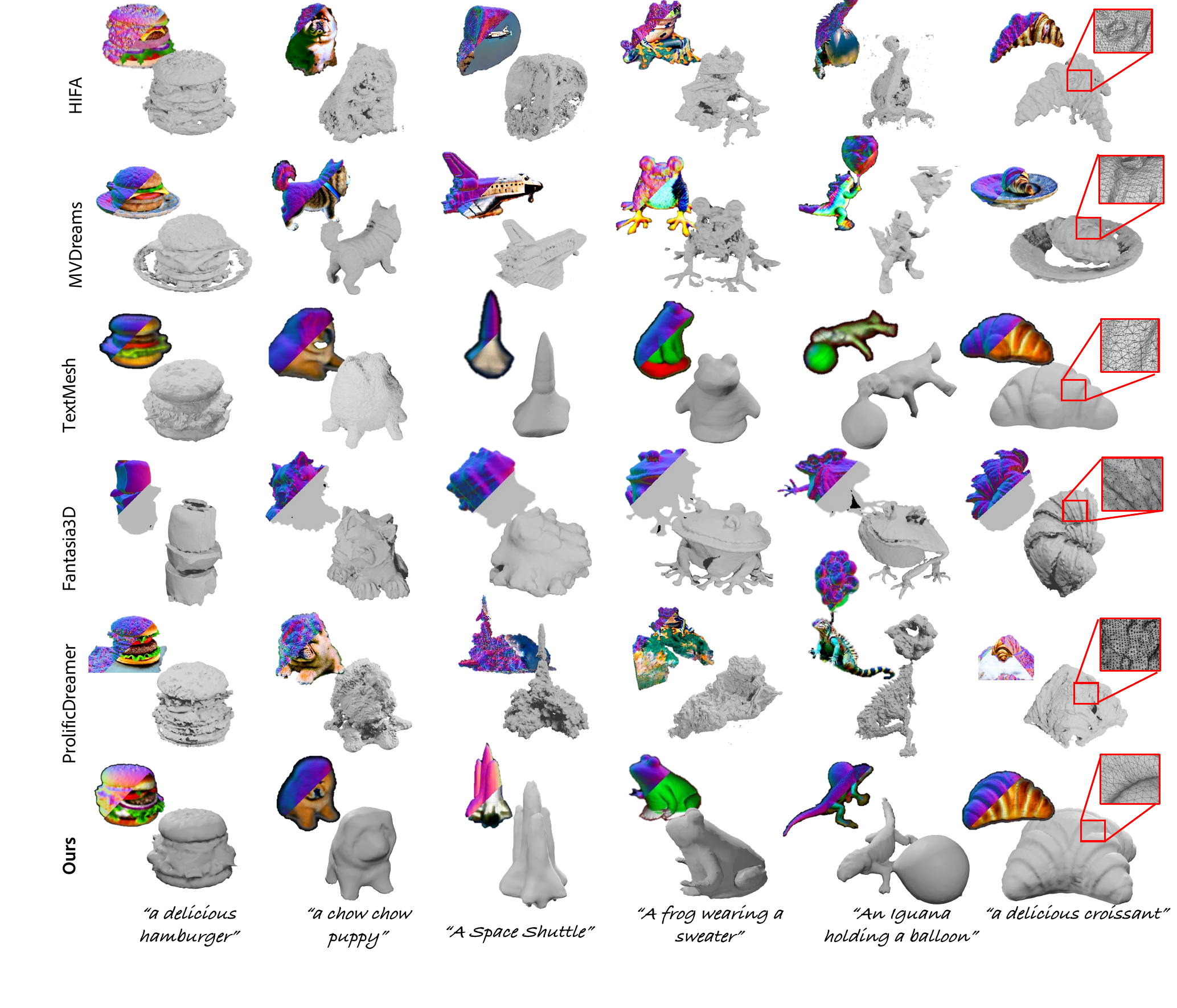}
    \caption{\tg{{\bf Qualitative comparison on text-to-3D  from scratch.} To validate the benefit of our hybrid approach, we compare the quality of the triangular meshes extracted from HIFA~\cite{zhu2023hifa}, MVDreams~\cite{shi2023mvdream}, TextMesh~\cite{tsalicoglou2023textmesh}, Fantasia3D~\cite{chen2023fantasia3d} and ProlificDreamer~\cite{wang2023prolificdreamer}. 
    For each prompt and method, we show the normal and RGB rendering on the top left, and the textureless mesh on the bottom right. While all methods produce realistic RGB renderings, only our hybrid representation generates smooth geometry, as highlighted by the red insets.}}
    \label{fig:gen_comparison}
\end{figure*}

Since \ourmethod{} is a modeling tool, we are primarily interested in evaluating the quality of the geometry and thus focus on mesh renderings without texture. Note that most existing 3D generative techniques are not designed to edit a part of an existing mesh, and therefore we start by comparing the performance of our hybrid approach for unconditional text-to-3D generation, seeking to establish the ability of our method to generate higher-quality meshes.
We compare against five recent approaches: HIFA~\cite{zhu2023hifa}, MVDream~\cite{shi2023mvdream}, Fantasia3D~\cite{chen2023fantasia3d}, ProlificDreamer~\cite{wang2023prolificdreamer} and TextMesh~\cite{tsalicoglou2023textmesh}. \tg{We run the open-source implementation of all methods to produce results. We use Marching Cube to extract a mesh from NeRF-based methods, and isosurface extraction for SDF-based methods. We do not compare against Instant3D~\cite{instant3d2023} because they do not provide an implementation as of writing this paper, but we note their ability to generate high-quality meshes.}
%
We measure the quality of the generated objects using an average clip score between the prompt and multi-view renderings. A complete description of our metric and the prompts used are given in the supplementary materials.
We present our results in Figure~\ref{fig:gen_comparison} \tg{and Table~\ref{tab:quantitative}}. 
Extracted meshes often exhibit significant surface artifacts, which make them hardly recognizable without texture (see ``Chow Chow puppy'' by  ProlificDreamer or ``Croissant'' by TextMesh). 
By comparison, our geometries are recognizable and smooth thanks to our hybrid representation enabling an explicit regularization of the surface. \tg{We further verify that, among competing approaches, \ourmethod{} achieves the highest Clip Score between the prompts and texture-less renderings of the mesh. 
This shows that \ourmethod{} achieves semantically meaningful reconstruction via geometry and not only texture, and thus successfully bridges the generative capabilities of implicit radiance fields with the surface-level controls of meshes.} 

\begin{figure}
    \centering
    \includegraphics[width=0.9\linewidth]{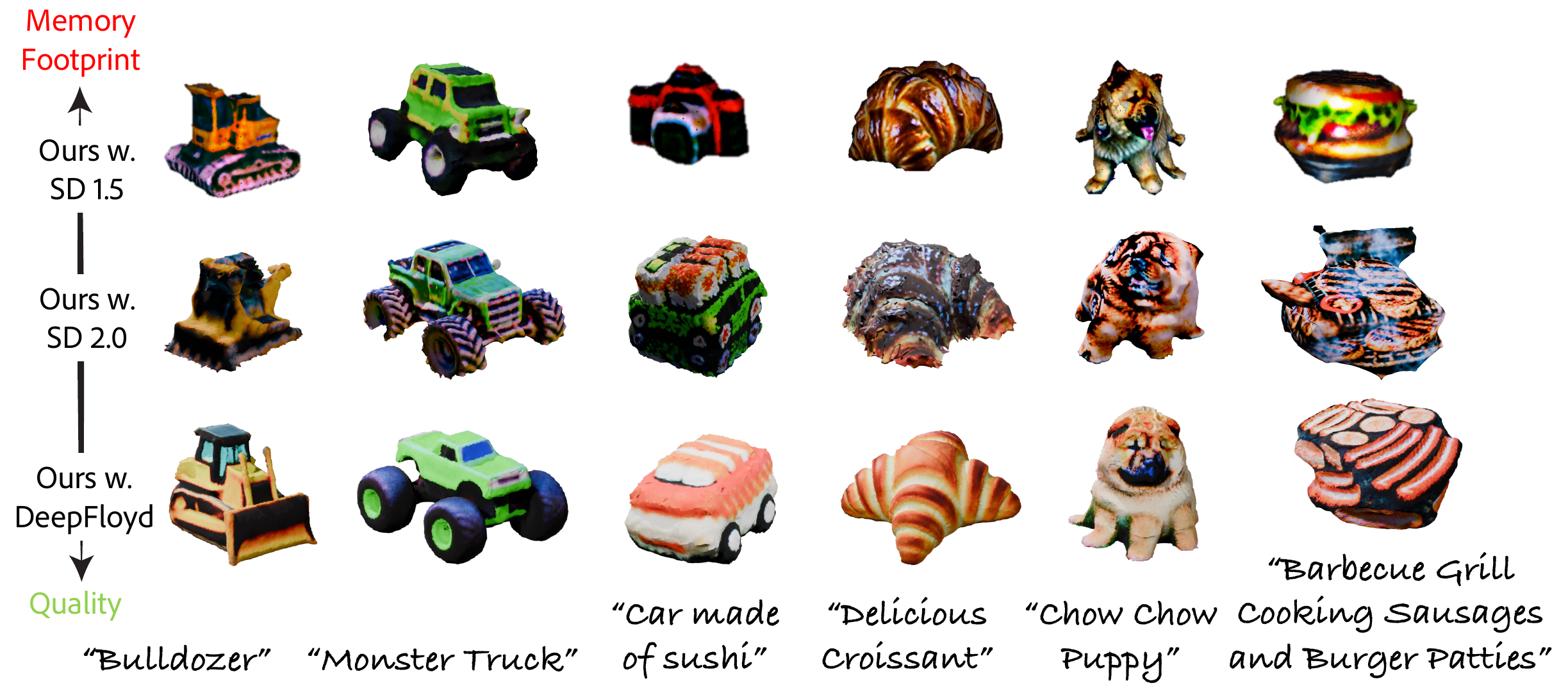}
    \caption{\tg{{\bf Varying the backbone of \ourmethod{}.} We show that \ourmethod{} is orthogonal to the backbone text-to-image model used to provide gradients in Score-Distillation Sampling. We observe a quality and required VRAM trade off for various models.}}
    \label{fig:orthogonality}
\end{figure}

\tg{
Our hybrid representation is orthogonal to the choice of backbone text-to-image models. Fig~\ref{fig:orthogonality} shows 3D models generated by \ourmethod{} with SD 1.5, SD 2.1, and DeepFloyd, showing that the backbone model influences the quality of the results. This drives two conclusions : (1) improvements in the backbone translate into improvements for 3D generation. (2) In addition to the representation, an important differentiating factor for each method in Table~\ref{tab:quantitative} is the backbone, which we report.
}

\subsection{Mesh Sculpting}
\label{subsec:sculpting}

 \ourmethod{} generates a modified mesh, which could be iteratively refined with new elements. Note that hybrid representation is essential to this application. Selecting the region of interest is easily accomplished by using standard mesh editing tools~\cite{blender}. Using the mesh allows us to keep non-selected surface regions intact by zeroing out their deformation gradients, which guarantees that the change will only affect the user-selected region. \tg{Because they are preserved, mesh properties can be transferred in the non-edited region, including texture, tessellation, and rigging parameters}. 
%
%
Refer to Figures~\ref{fig:teaser} for example results. \ourmethod{} generates high-quality edits that match the rough local edit and adhere to the user's text prompt. 

\subsection{Comparison with Contemporary 3D Editing Pipelines}
\label{subsec:sculpting_comparison}
\tg{A naive alternative to localized mesh sculpting would be to use the existing text-driven mesh deformation technique~\cite{gao2023textdeformer} and to zero-out the deformation field outside of the editing region, which is possible because the representation is explicit. We further compare with Latent-Nerf~\cite{metzer2022latent} and Vox-E~\cite{sella2023voxe}, but these approaches cannot guarantee a local edit because they are based on soft attention. We discuss InstructNeRF2NeRF~\cite{instructnerf2023} and DreamEditor~\cite{zhuang2023dreameditor} in supplementary} 

In Figure~\ref{fig:comparison}, we compare our method to these three baselines. Note how our method is able to add geometrically complex large-scale details due to guidance from SDF and topological updates. Our method's changes are also restricted only to user-selected regions and do not lead to any changes in the other parts of the input. 
%

\subsection{Analysis of Mesh and SDF Robustness to Noise}
\label{subsec:analysis}

\begin{figure}[t]
    \centering
    \includegraphics[width=\linewidth]{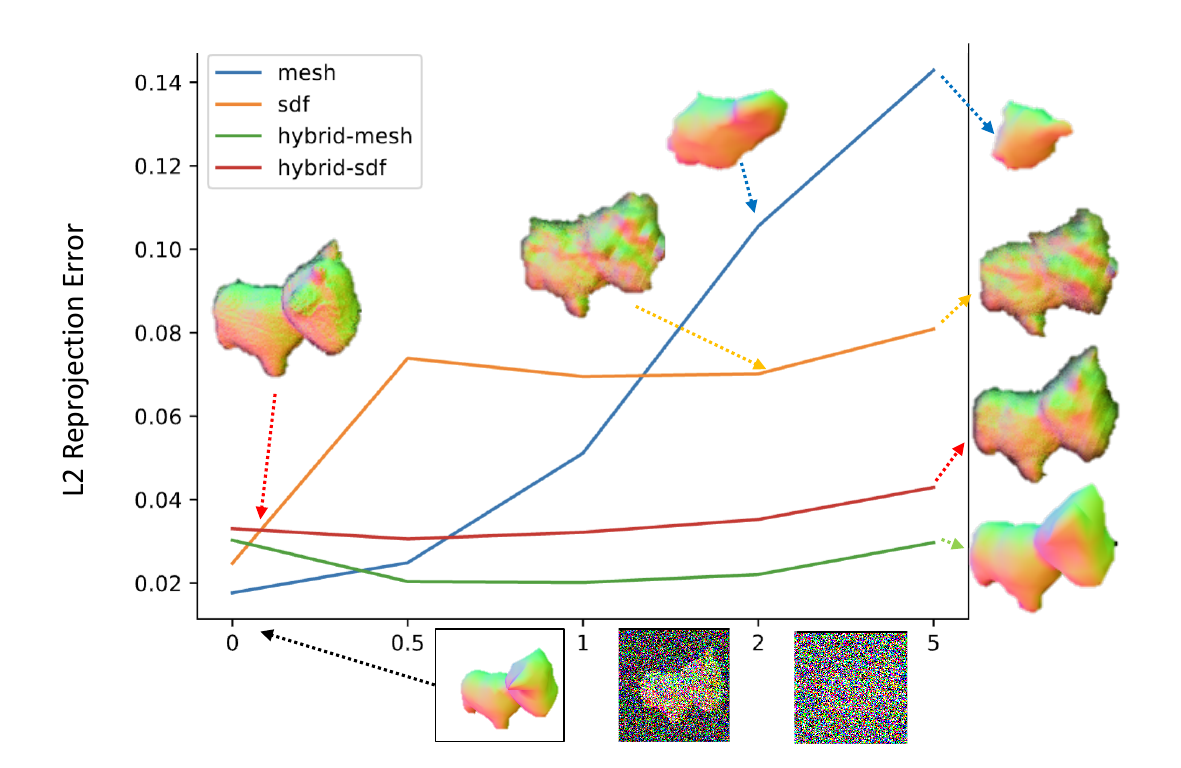}
    \caption{\bf{Mesh and SDF robustness to noisy gradients.} \textmd{We optimize a \textcolor{blue}{mesh}, an \textcolor{orange}{SDF} and our hybrid representation with multi-view reconstruction losses after applying various noise levels to the ground truth renderings. We report the L2 reprojection error against novel-views ground truth renders.  The SDF exhibits more robustness than the mesh to the high noise regime. We show the results for both the mesh (\textcolor{darkgreen}{hybrid-mesh}) and SDF (\textcolor{red}{hybrid-SDF}) in our hybrid representation. The \textcolor{darkgreen}{hybrid-mesh} significantly outperforms the \textcolor{blue}{mesh} only baseline in the high noise regime. 
    }}
    \label{fig:noiserobustness}

\end{figure}


We now motivate the use of our hybrid representation by a simple controlled experiment, where we aim to reconstruct a fixed 3D target with different levels of noise in the guidance. Even though we use synthetic noise, we expect these findings to apply in an SDS setting, where gradients are noisy due to the noising step performed at each SDS iteration~\cite{poole2022dreamfusion}. 


Given multi-view renderings of a fixed ground truth 3D model, we add uncorrelated per-pixel Gaussian noise to each image and compute L2 pixel-wise loss to guide our shape representation towards the target. A complete overview of the experiment setup is given in the supplementary materials. As we increase the noise level (by increasing standard deviation) we find that different representations are more prone to errors in reconstructing the target. We use L2 re-projection error with respect to the ground truth shape on novel views as our evaluation metric, and compare vanilla {Mesh}, {SDF} representations to our hybrid approach (using both {SDF} and {mesh} renderings), and show results in Figure~\ref{fig:noiserobustness}. 
\tg{
At the highest noise regimes (standard deviation of 5), the {mesh reconstruction} degenerates to a blob, while the {SDF reconstruction} is still recognizable despite surface irregularities. Importantly, the hybrid representation performs better than both individual representations at all levels of noise, and the benefits are the strongest at higher noise levels. {The hybrid-SDF} outperforms the {SDF baseline} because the mesh acts as a regularizer for more 3D consistency.  The L2 reprojection loss of the {hybrid-mesh} is lower than the {hybrid-SDF} because the SDF tends to exhibit a “haze” artifacts throughout the entire image, including the background pixels, while the mesh rendering can not create this effect in the background, serving as a sort of regularization on its own. Finally, we note that all curves are not strictly increasing, which we attribute to the fact that we test on novel views: some amount of noise may prevent overfitting to the training views.

Note that this experiment is not designed to claim superiority of the hybrid representation over the {SDF}, but rather to show its robustness to noise.
}
\subsection{Ablations}
\label{subsec:ablations}

\begin{figure}
    \centering
    \includegraphics[width=0.89\linewidth]{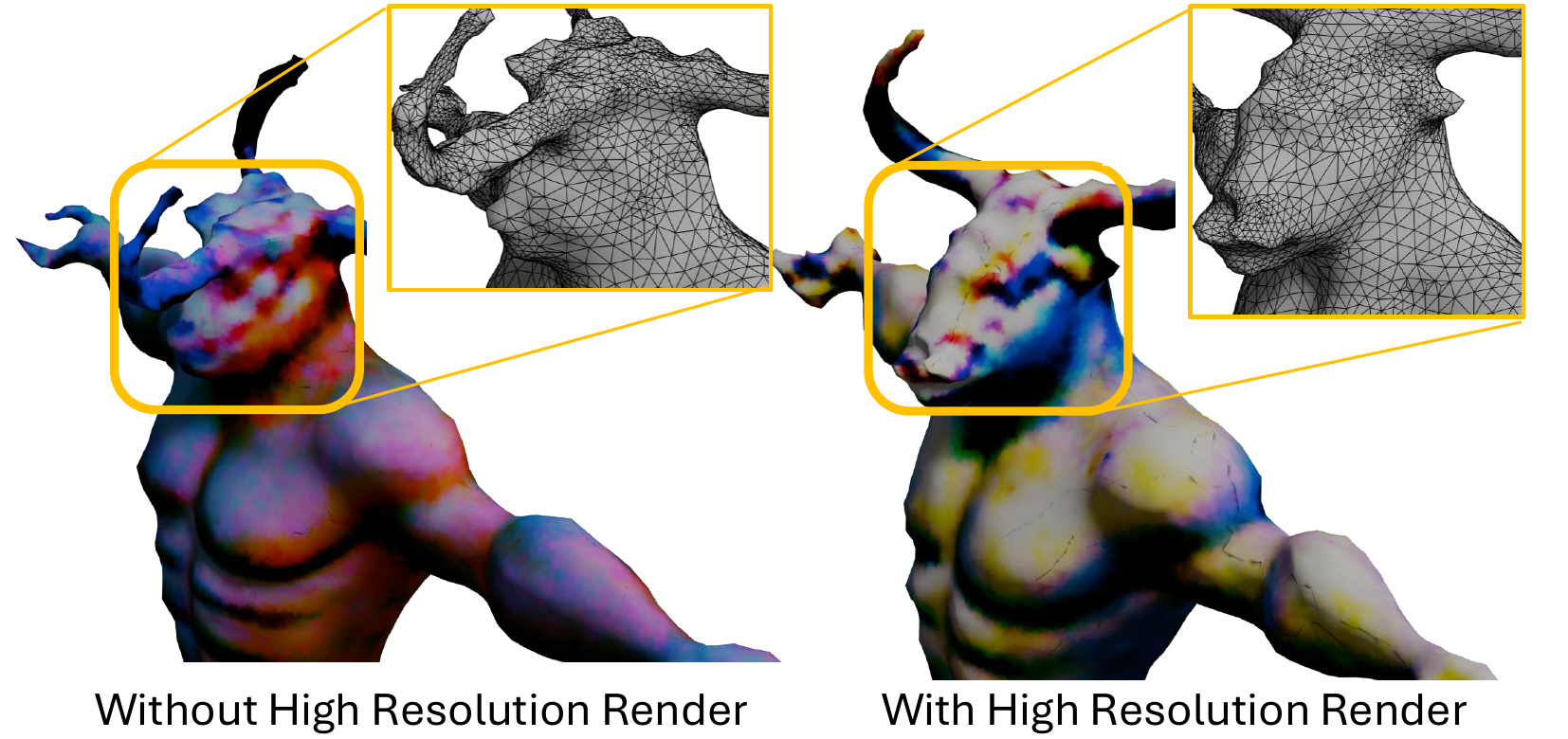}
    \label{fig:highres}
    \caption{\tg{{\bf Ablation on high resolution renderings.} \textmd{Without our scheme to render SDF in high resolution by using the mesh counter part to localize samples along the ray, we default to regular low resolution SDS on the SDF renderings, leading to poorer quality in the generated shapes.
    }}}
    \label{fig:ablation-no-highres}
\end{figure}

\paragraph{512x512 SDF renderings} \tg{We show in Figure~\ref{fig:ablation-no-highres} and Table~\ref{tab:quantitative} that high-resolution SDF rendering significantly affects generated shape quality. This shows that the mesh part of the hybrid helping to accelerate SDF rendering plays a role in quality.}

\begin{figure}
    \centering
    \includegraphics[width=0.55\linewidth]{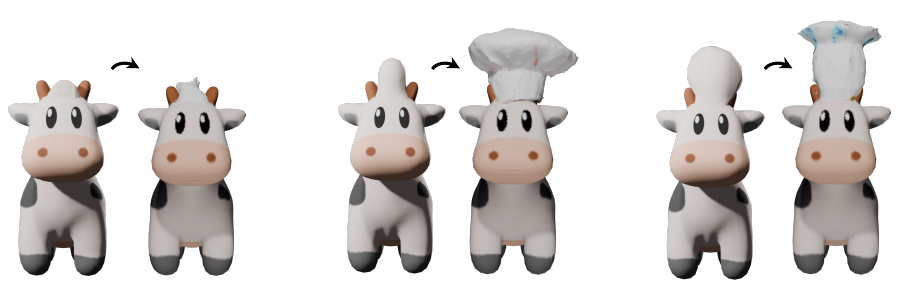}
    \caption{{\bf Robustness to the initial edit.} \ourmethod{} sculpts the hat from various levels of manual initalization. We find that similar initial edits give similar results (middle and right). No edit or small edits (left) lead to smaller changes, after the allocated 10k iterations. \amir{Note that \ourmethod{} is \textit{not} invariant w.r.t. the initial edit, as it is not desirable in a sculpting workflow.}
}

    \label{fig:intialization}
\end{figure}

\paragraph{Initial editing. }
\tg{In our experiments, 
only significant differences in the initial sculpt lead to different generated results. Typically, performing an initial sculpt does impact the result compared to just selecting a region with no sculpting. However, two initial sculpts that are similar will lead to the same final results (see Figure~\ref{fig:intialization}).}

\begin{figure}
    \centering
    \includegraphics[width=0.73\linewidth]{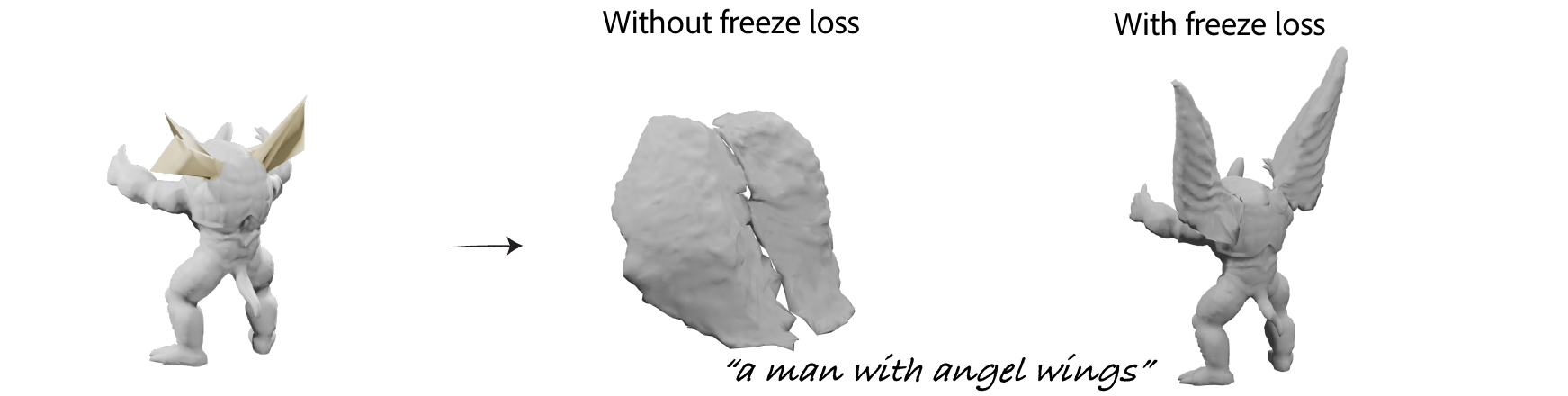}
    \label{fig: freeze loss}
    \caption{\tg{{\bf Ablation: no localization.} \textmd{W/o localization and freeze losses,  shape changes can propagate beyond the user-selected area, potentially destroying the initial content. Here, armadillo is erased by ``angel wings.''
    }}}
    \label{fig:ablation-no-freeze}
\end{figure}

\paragraph{Not enforcing localization, no freeze loss}
We remove the mechanism for enforcing localization via fixing non-selected surface regions and nearby SDF values as discussed in Sec.~\ref{subsec:priors}. In Figure~\ref{fig:ablation-no-freeze}, The shapes undergo unintended global changes, potentially erasing the original shape.

\begin{figure}
    \centering
    \includegraphics[width=0.89\linewidth]{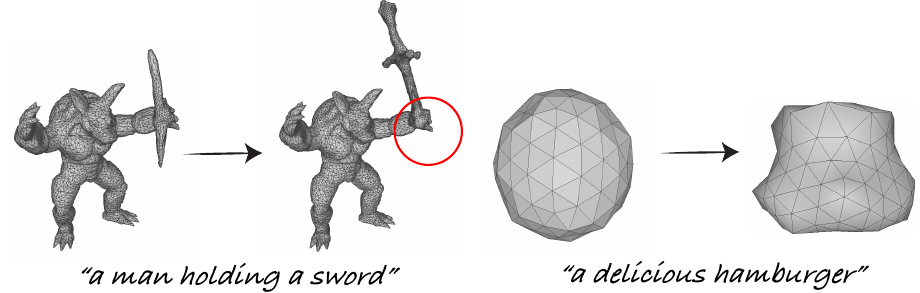}
    \caption{\tg{{\bf Ablation: no topology updates.} Optimizing the mesh without topology update results in the final generated object being limited by the initial resolution. \textbf{Left} When starting with a fine mesh the optimization will often get stuck since each vertex has tiny effect on the objective: notice the sword is unable to grow its tip. \textbf{Right} When starting from a coarse mesh, no fine details can be created.}}
    \label{fig:ablation-no-topology}
\end{figure}

\paragraph{No topology updates}
The topological updates (Sec.~\ref{subsec:topology}) allow to add resolution gradually. Optimizing a fixed-resolution mesh would either result in a shape that only marginally differs from the input if the initial resolution is too high or lacks fine details if the initial resolution is too coarse (Figure~\ref{fig:ablation-no-topology}).
\tg{These results complement, in a generative setting, the  experiments of \cite{continous_remshing2022}\footnote{We refer the reader to the videos in the Readme of their GitHub repository}, performed with ground truth multi-view supervision which similarly shows that optimizing very high-poly meshes leads to local minima.}

\section{Conclusion}
\label{sec:conclusion}

We presented \ourmethod{}, a generative sculpting tool backed by our new hybrid SDF and mesh representation. Key to the success of the generative process is our new rendering strategy that leverages the mesh part of the hybrid representation to localize ray sampled in the volumetric rendering of the SDF. We believe \ourmethod{} is an important step towards turning the recent advancements in text-to-image-from-scratch into an actual modeling tool usable by artists in an iterative workflow.

\paragraph{Limitations} Our method is inherently constrained by the quality of the SDS gradients. 
Each view tracts the optimization in a different direction adding noise and rendering the emergence of fine details more difficult. 
Second, \ourmethod{} is not interactive, \eg running ~\ourmethod{} takes 1 hour per prompt on an A100 GPU, with\tg{The bottleneck being the SDS loss}


\paragraph{Future work} 
We see opportunities to leverage inpainting and depth-conditioned diffusion models to speed up the convergence of SDS. Indeed, this generative process transforms the full object in each rendering, whereas it is clear that some parts of the generated image should stay the same as the 3D edit is localized. We think that leveraging this insight would reduce the amount of noise inherent in SDS.
%


\bibliographystyle{ACM-Reference-Format}
\bibliography{sample-base}

\begin{figure*}
    \centering
    \includegraphics[width=0.85\textwidth]{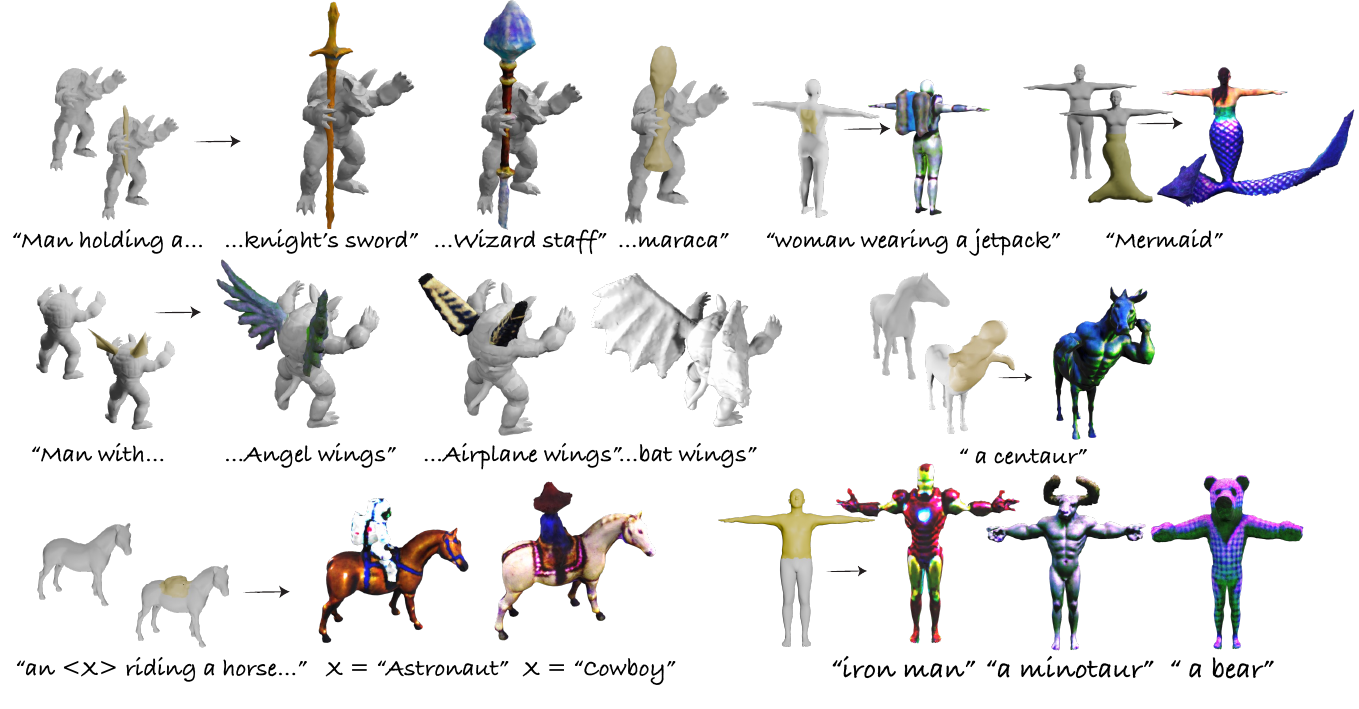}
    \caption{{\bf Sculpting gallery.} \textit{Left:} \textmd{from a source mesh, the user performs a rough edit in under two minutes, highlighted in yellow. \textit{Right:} \ourmethod{}'s output.}}
    \label{fig:gallery-text-based}
\end{figure*}

\begin{figure}
    \centering
    \includegraphics[width=0.85\linewidth]{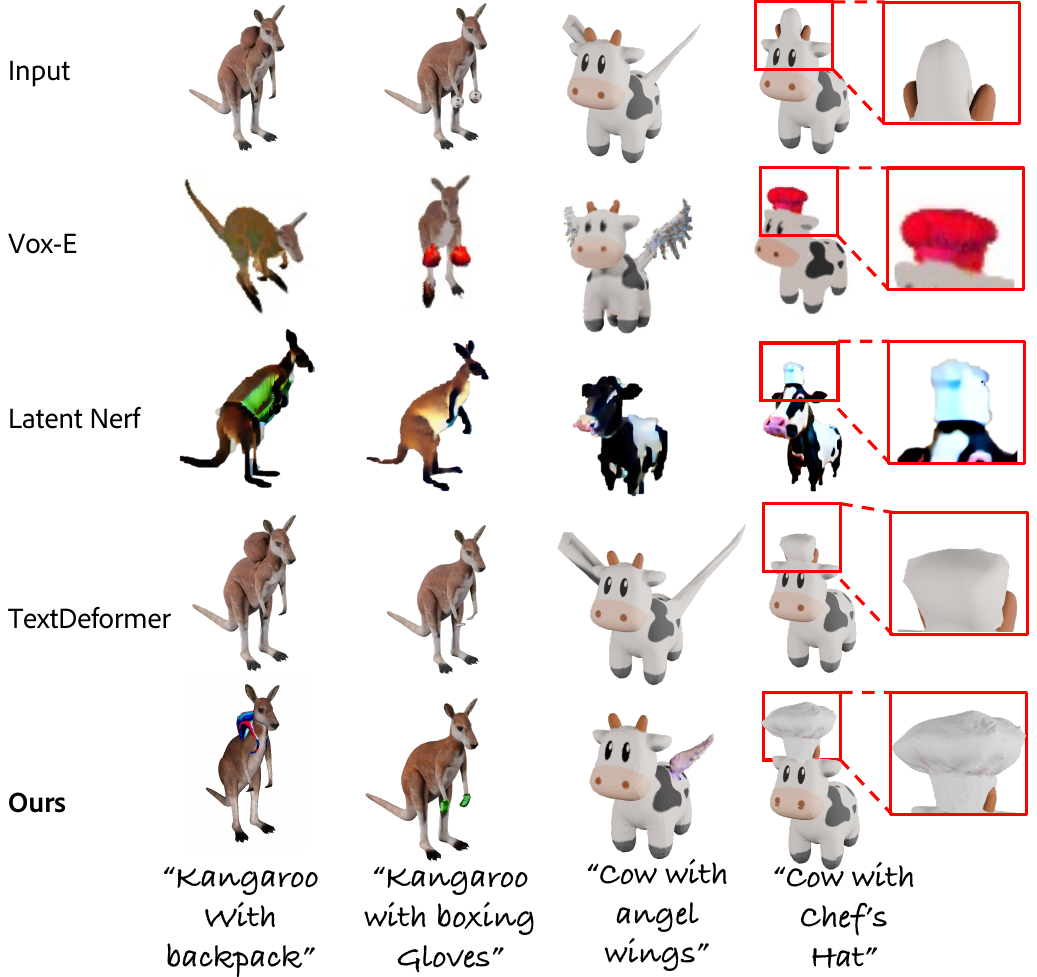}
    \caption{{\bf Comparisons to other 3D-editing approaches}. We compare \ourmethod{} with other generative approaches able to edit meshes~\cite{sella2023voxe, metzer2022latent}. Note that the baselines \textit{do not} strictly preserve the non-editing region and texture. They rely on soft attention, which leads in some cases to the inadvertent destruction of features such as the cow's horns. In contrast, our method is guaranteed to be non destructive for the existing geometry and preserves its texture. We post-process TextDeformer~\cite{gao2023textdeformer} to support localized edits, and show that the space of deformation it can achieve is less expressive.}
  
    \label{fig:comparison}
\end{figure}

\clearpage

\end{document}


\maketitle

\hspace{10mm}

\section {Implementation Details}
For reproducibility, we describe the settings used during the optimization. We intend to release the code upon acceptance.

\subsection{Initialization} 
\begin{figure}[h]
     \centering
         \includegraphics[width=0.8\linewidth]{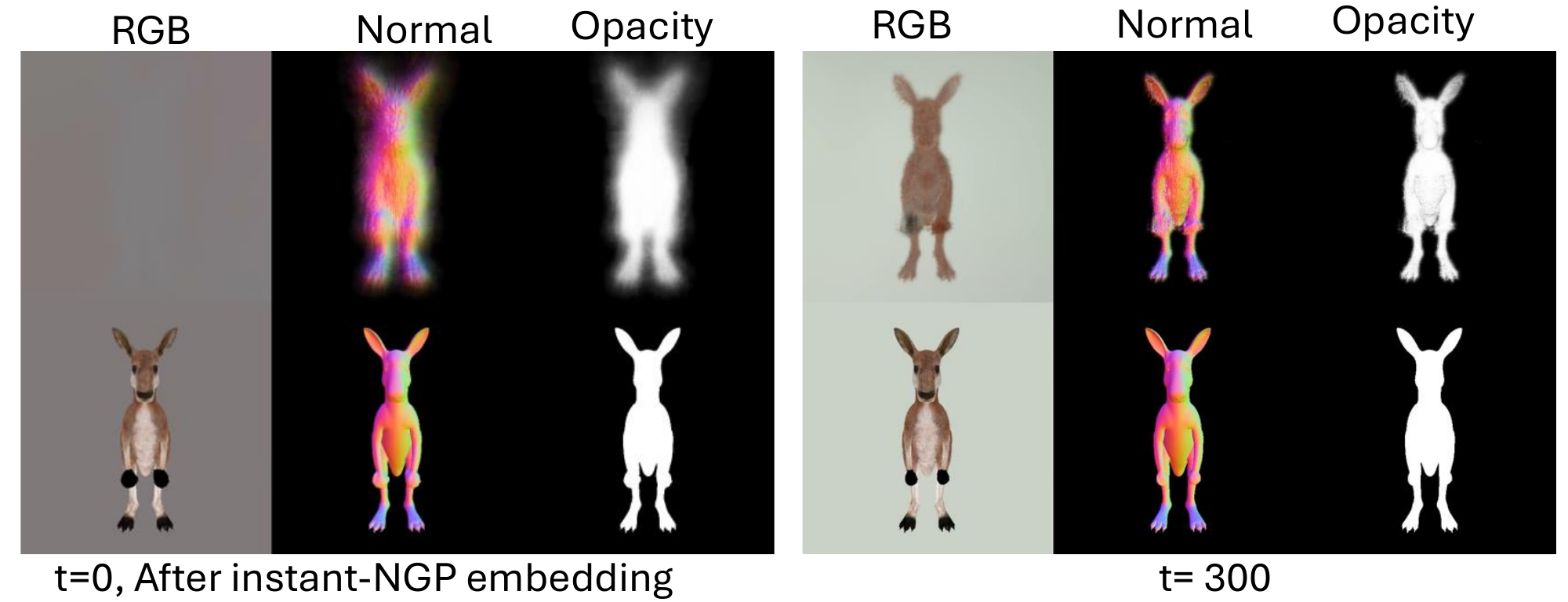}
        \caption{\textbf{Effect of the two-stage initialization.} \textit{Top row}: SDF render. \textit{Bottom row}:  mesh render. L2 multiview consistency losses provides a more consistent starting point for the hybrid representation, in addition to allowing initializing the SDF appearance network based on the initial mesh texture on non-editable areas.}
        \label{fig:initialization}
\end{figure}

To initialize our hybrid representation at a consistent state, the mesh is encoded as an SDF at the start of the optimization. We use the pySDF library to get ground truth SDF values of the mesh, and in each iteration sample 1M points as described in \cite{instantngp}, with the loss being the sum of the square differences between each samples value evaluated with the SDF network and the ground truth SDF.  Due to the fact that the SDF rendering is sample based, the resulting SDF renders are usually blurry, as seen in Figure~\ref{fig:initialization} Stage 1. The sampling done along the rays is dynamic using the NeRFACC library~\cite{li2023nerfacc}. To correct the blurriness and get a more consistent starting point for the optimization (for example, the one shown in Figure \ref{fig:initialization} Stage 2), we further optimize L2 multiview consistency losses for 300 iterations.

\amir{
\subsection{SDF representation} The SDF network uses hashgrid encodings similar to Instant-NGP~\cite{instantngp}, which allows to significantly shrink the MLP and consequently the cost of SDF queries, making the initialization take ~5-10 minutes for 2000 iterations on a A100 GPU.
We use the TextMesh ~\cite{tsalicoglou2023textmesh} implementation in ThreeStudio ~\cite{threestudio} as our SDF backbone, with the default hyper-parameters (namely, $\lambda_{Eikonal}$ = 1000).

\subsection{Training Procedure} We run our pipeline for 10000 iterations. and use $\lambda_{Normal}$ = 0.1 and $\lambda_{RGB}$ = 0.1 and $\lambda_{mesh_smoothness}$ = 0.1.
Additionally, we apply the mesh smoothness loss only to non-frozen vertices.

\subsection{mesh learning rate schedule}
    We use ~\cite{barda2023roar} as the re-meshing backbone. We set the learning rate to be 0.1, which we found provides a good balance between mesh responsiveness to the current SDF state while still keeping a stable and uniform expansion front. for the last 500 iterations we reduce the learning rate to 0.05, to promote convergence to a smoother mesh.
}

\subsection{Topology updates to maintain the border between editable region and non-editable region}
Special consideration must be given to topology edits when dealing with edges or faces that have a mix of frozen (i.e, not editable) and editable vertices. We outline the rules used in our pipeline:
\begin{enumerate}
    \item \textit{Face split.} If one or two of the face's vertices are frozen this face can still be split, and the newly created vertex is editable
    \item \textit{Edge collapse.} If one of the edge's vertices is frozen the edge may be collapsed, with the new vertex belonging to the frozen set of vertices.
    \item \textit{Edge split.} For each edge, we observe it's one-ring vertices. An edge may be flipped if one of these two conditions are met: (i) any of its opposite vertices are editable or (ii) both of its vertices are editable.
\end{enumerate}



\section{Evaluation}

\subsection{Unconditional text-to-3D}

\begin{figure}[h]
     \centering
         \includegraphics[width=0.8\linewidth]{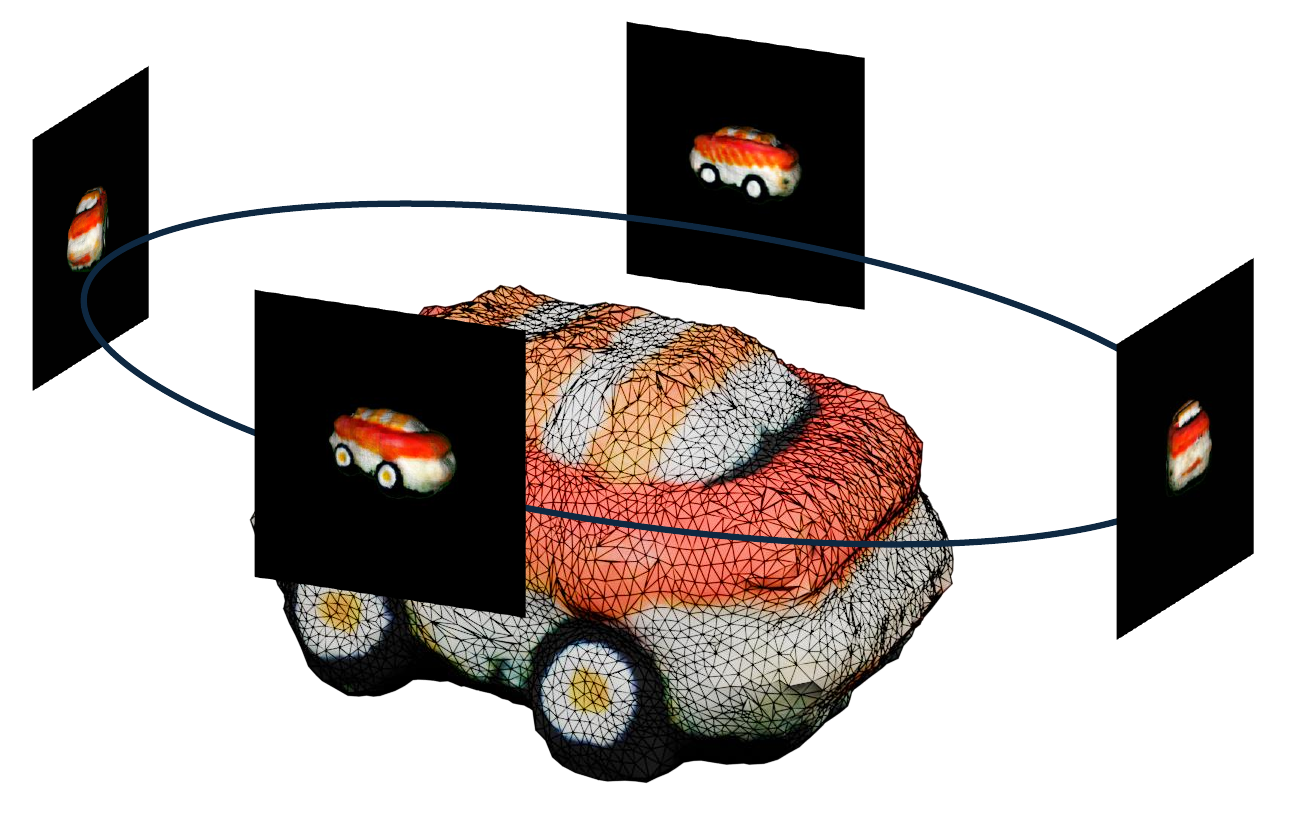}
        \caption{\textbf{Clip Score calculation.} We illustrate four cameras out of 20 that are used to render the mesh and compute a averaged Clip score.}
        \label{fig:avg_clip_score}
\end{figure}

\paragraph{Average Clip Score Calculation}
\amir{
We average the clip score over 20 renders uniformly over a circular camera pattern, illustrated in Figure~ \ref{fig:avg_clip_score}.
The camera parameters used (using the standard Threestudio \cite{threestudio} camera definitions)
\\
$
elevation\_deg = 15^o \\
elevation\_deg = (-10, 90) \\
azimuth\_range = (-180,180) \\
camera\_distance = 3.0 \\
eval\_fovy\_deg = 70^o \\
$
}

\paragraph{List of prompt in the benchmark}
 \amir{The prompts are randomly chosen from the Threestudio curated list.} 
\begin{enumerate}
\item A DSLR photo of a Space Shuttle
\item A DSLR photo of a pug wearing a bee costume
\item A DSLR photo of a roast turkey on a platter
\item A DSLR photo of a squirrel made out of fruit
\item A DSLR photo of a toy robot
\item A DSLR photo of a tree stump with an axe buried in it
\item Flower made out of metal
\item A lionfish
\item A plush dragon toy
\item A typewriter
\item A DSLR photo of a barbecue grill cooking sausages and burger patties
\item A DSLR photo of a bulldozer
\item A DSLR photo of a car made out of sushi
\item A DSLR photo of a chow chow puppy
\item A DSLR photo of a delicious croissant
\item A DSLR photo of a frog wearing a sweater
\item A DSLR photo of a ghost eating a hamburger
\item A DSLR photo of a green monster truck
\item A DSLR photo of an iguana holding a balloon
\item A delicious hamburger
\end{enumerate}

\subsection{More baselines for local generative editing.}
\begin{figure}[h!]
    \centering
    \includegraphics[width=\linewidth]{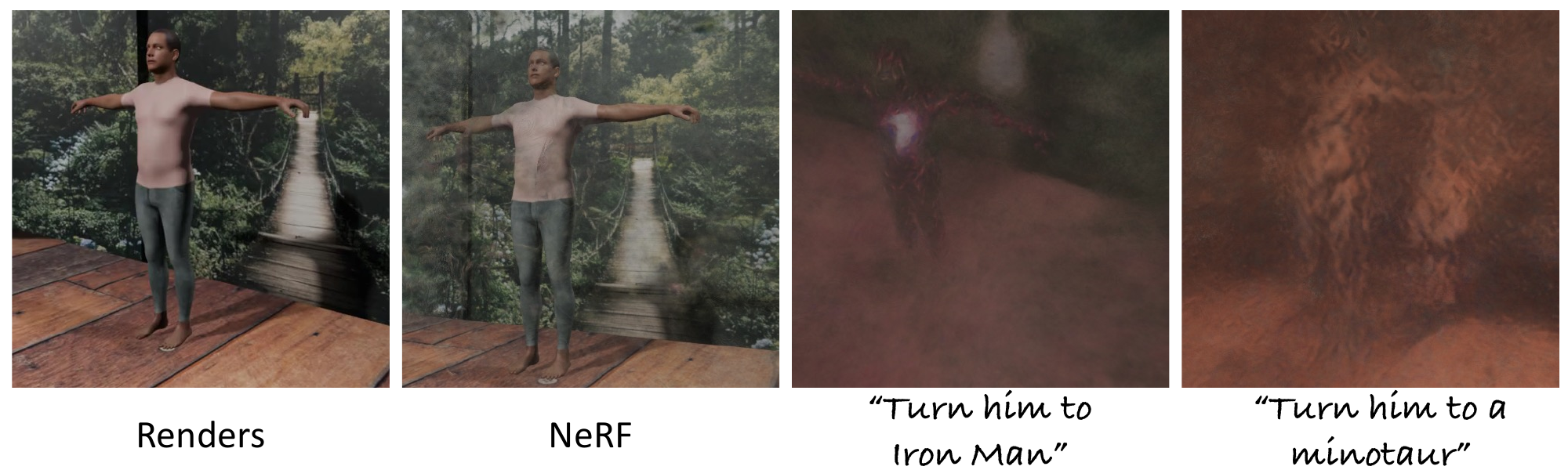}
    \vspace*{-8mm}
    \caption{IntructNerf2Nerf results on synthetic data. \textmd{IN2N uses NeRFStudio's Instruct-Pix2Pix as a backbone, which works best on real image. We show our attempts to use IN2N for synthetic data (male human mesh from the SMPL dataset). Applying IN2N on the resulting NeRFs lead to very blurry results.}}. 
    \label{fig:in2n_image}
\end{figure}

\paragraph {Instruct Nerf2Nerf Results}
InstructNerf2Nerf~\cite{instructnerf2023} is a recent pipeline for editing NeRF scenes using prompts. The official implementation uses Instruct-Pix2pix~\cite{brooks2022instructpix2pix}, which is trained on real images and is not suitable for synthetic data, Therefore it does not work for sculpting an existing 3D mesh because of the domain gap. Our best attempts at applying Instruct-Nerf2Nerf on synthetic data are given in fig \ref{fig:in2n_image}. An email has been sent to the author, but as of the submission date, no response has been received.

\begin{figure}
    \centering
    \includegraphics[width=\linewidth]{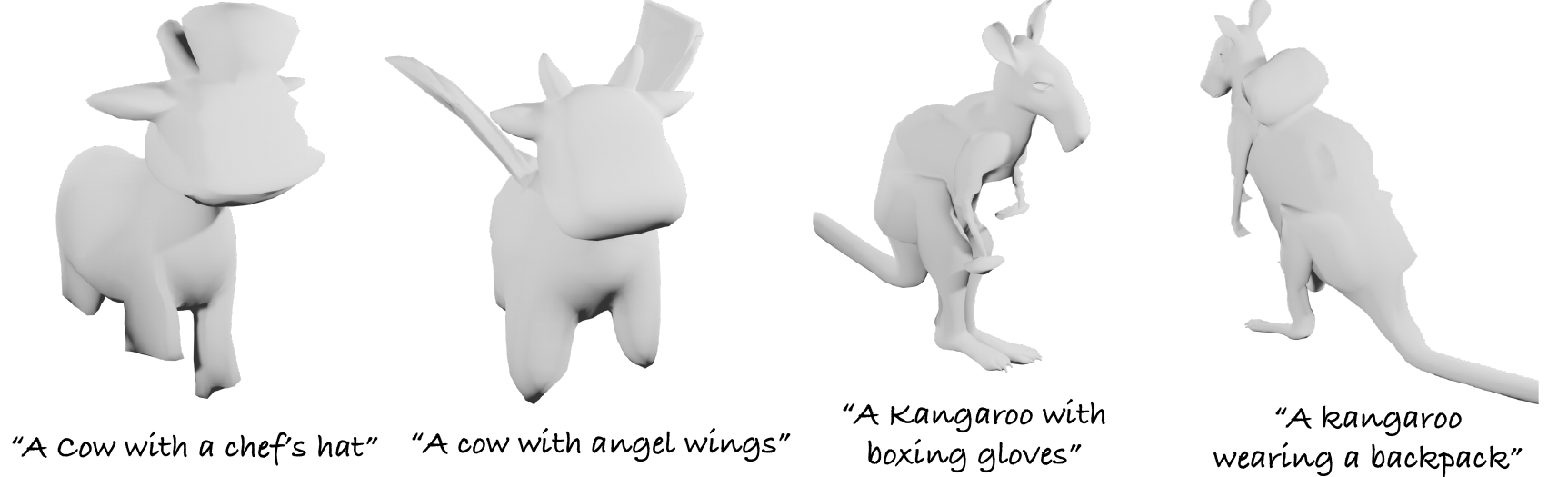}
    \caption{\textbf{TextDeformer output without post-processing}. To complement the results presented in the main paper, we show the results of TextDeformer without our post-processing to localize the edit. As can be seen, the shape is globally affected. 
    }. 
    \label{fig:textdeformer results}
\end{figure}

\paragraph{TextDeformer details}
Naively, one can attempt to perform editing using TextDeformer~\cite{gao2023textdeformer} by simply masking the gradients for all frozen vertices. In practice, TextDeformer does not perform the optimization directly on the vertices, but rather on their Jacobians, making decoupling the gradients for individual vertices not trivial, which in our experiments resulted in failure to converge. In order to apply TextDeformer for mesh editing tasks, we perform a simple post process on the results: we first run TextDeformer on the input mesh with the input prompt to get the results shown in Figure~\ref{fig:textdeformer results}. 
We combine the meshes by taking the frozen vertices of the original mesh and the non-frozen vertices of the edited mesh.

\paragraph{DreamEditor}
The official implementation for DreamEditor~\cite{zhuang2023dreameditor} has a demo that runs on the authors' preprocessed data. However, running the method on new data is significantly challenging. Other users have opened issues on Github and the matter remains open as of submission time.

\subsection {Further Details on Reconstruction Experiment}
\begin{figure}[b]
    \centering
    \includegraphics[width=\linewidth]{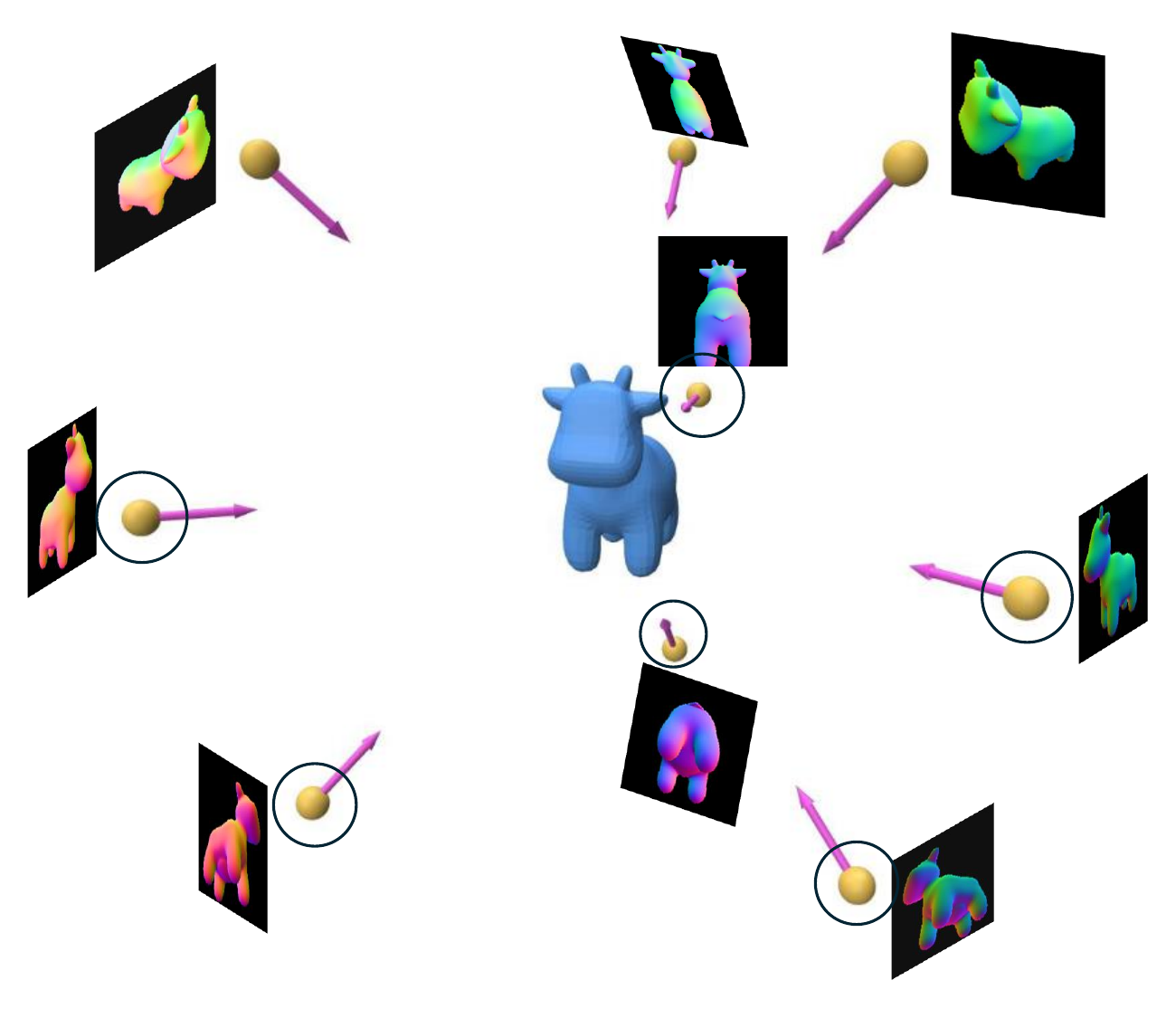}
    \caption{Camera setup for the multi-view reconstruction experiment. \textmd{The cameras are set up in a 3X3 configuration uniform on the unit sphere, camera for training views are circled.}}. 
    \label{fig:reconstruction_setup}
\end{figure}

The model used was "Spot" \cite{crane2013robust} . The cameras were spread uniformly on the unit sphere in a 3X3 grid going over the elevation and azimuth axis respectively, as shown in Figure~ \ref{fig:reconstruction_setup}, producing 9 different renders.

\subsection{Further ablation}
\paragraph{No face super-sampling for mesh colors}
\begin{figure}
    \centering
    \includegraphics[width=\linewidth]{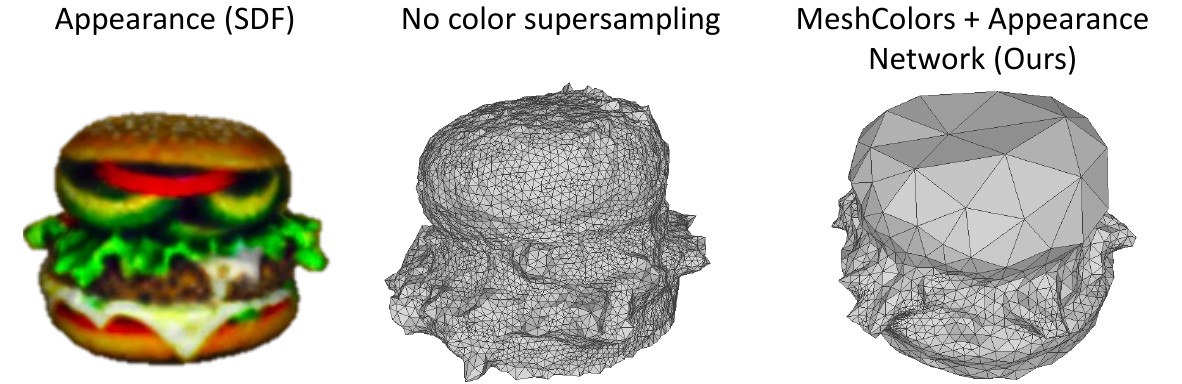}
    \label{fig: freeze loss}
    \caption{{\bf Ablation: no color supersampling.} \textmd{
    Using mesh-guided supersampling in conjunction with the appearance network allows to decouple geometry and appearance. Using this approach (top right), large faces are used for the mesh, even though the rendering still presents high frequency colors (bottom). When using a color-per-vertex scheme (top middle), significantly larger mesh resolution is required to achieve similar appearance.}}
    \label{fig:ablation-no-supersampling}
\end{figure}
Figure~\ref{fig:ablation-no-supersampling} illustrates the need for the mesh-driven super-sampling of the appearance network based on Mesh Colors scheme explain in the Method section of the main paper. We optimize our representation with respect to a target image (first column), either sampling colors per-vertex (second column), or using our adaptive sampling using MeshColors (third column). Note that without MeshColors sampling a much higher resolution is needed, which leads to poorer reconstruction and an over-tesellated mesh.

\section{Application}

\subsection{Application: Animation Transfers}
In production, meshes typically carry animation information. For instance, a rig \ie a collection of joints connected in a tree graph structure. Each joint is coupled to a group of vertices, so rig movements can be translated to mesh movement. This coupling is given by vertex groups, which is part of the mesh metadata. Because the non-editing regions are preserved, \ourmethod{} allows us to carry over this information through the optimization.

\begin{figure}
    \centering
    \includegraphics[width=\linewidth]{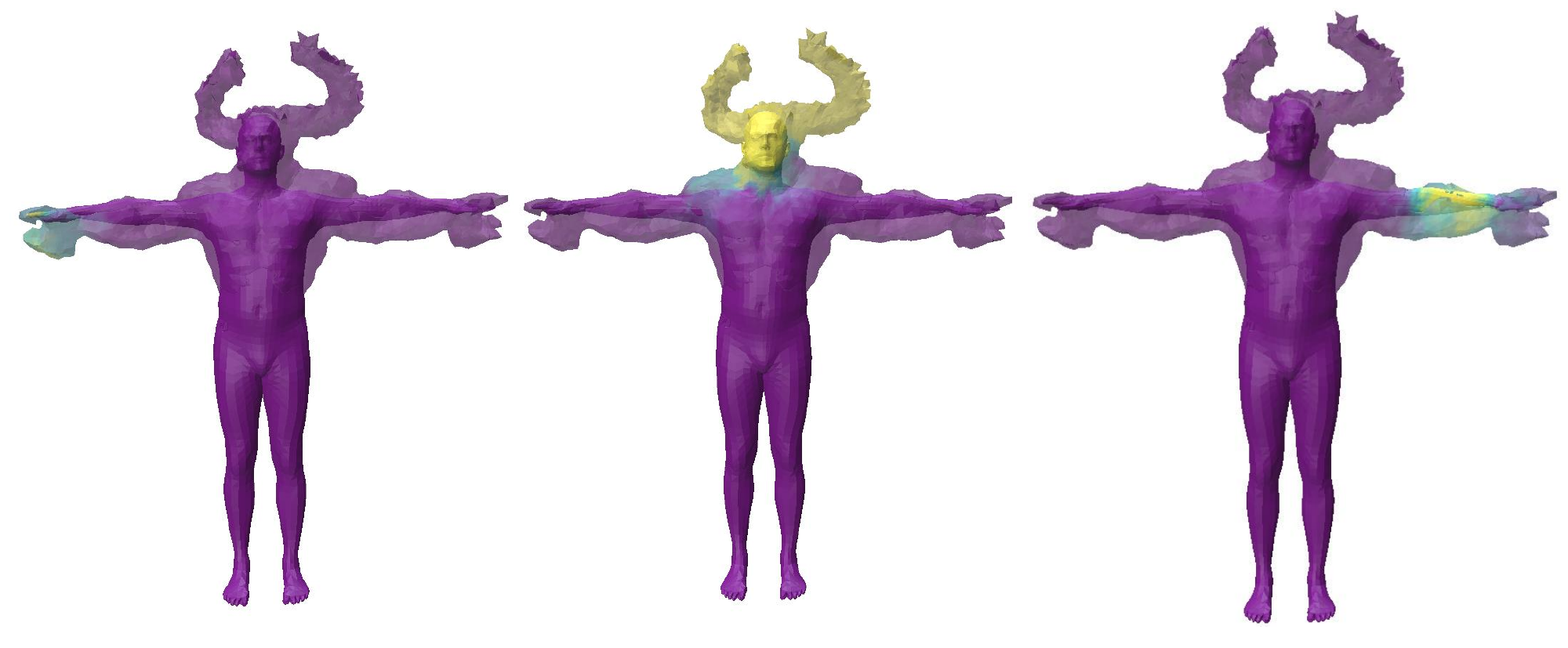}
    \includegraphics[width=\linewidth]{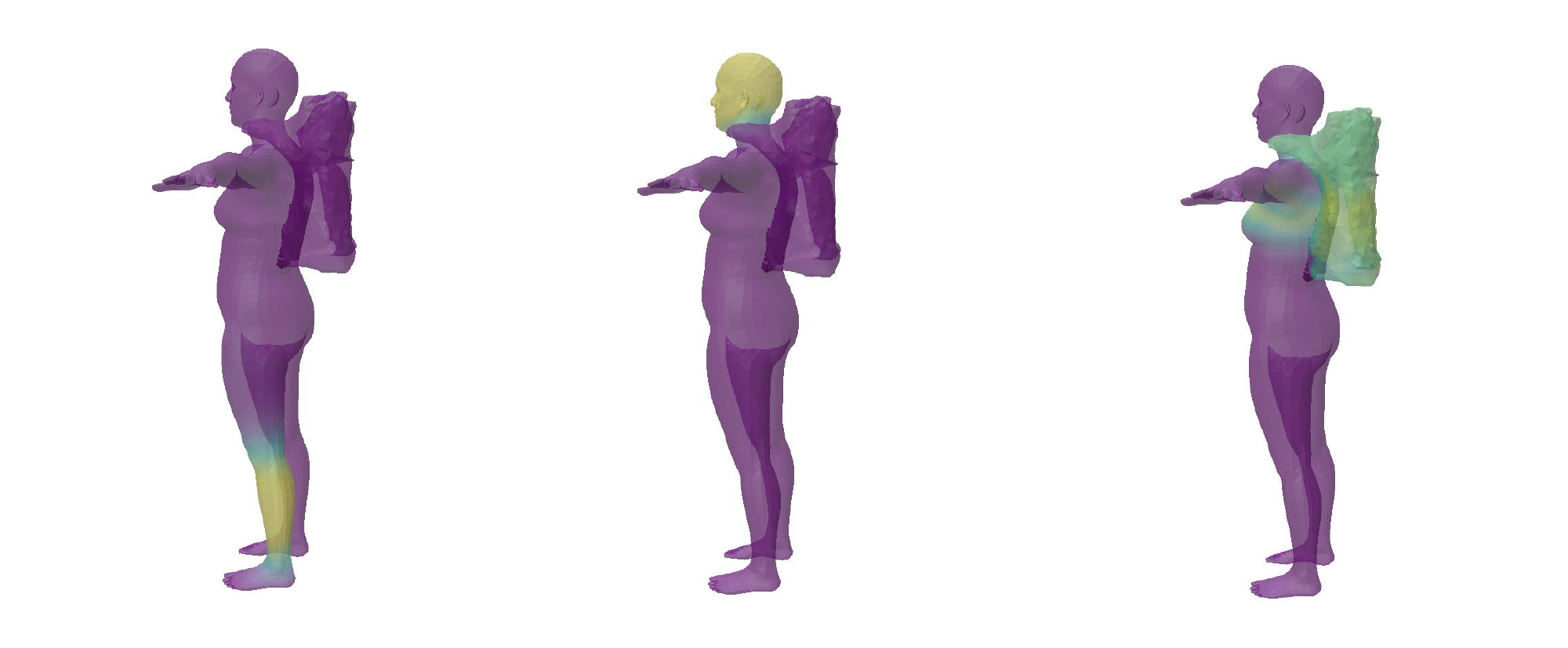}
    \caption{\bf{Vertex Group Transfer.} \textmd{ We superpose the mesh before and after editing by \ourmethod{}. We color the vertices based on their membership to a specific vertex group. In each column, we choose a different vertex group. The topology updates of ROAR~\cite{barda2023roar} preserve vertex features, such as the vertex group used in animations.}}
    \label{fig:animation}
\end{figure}

Vertex group assignment are trivially preserved in the non-editing region. For the editing region, we simply need to ensure that the information is preserved through the various topology updates of ROAR~\cite{barda2023roar}. For each vertex, we add a vector of size \#VG, which is a one-hot encoding for the respective vertex group. For face splits, we compute a weighted average of the neighboring vertices and take the maximum to assign the new vertex to a group. If an edge is collapsed, we pick the one-hot encoding of the adjacent vertex with the highest Q-slim score. As can be seen in Figure~\ref{fig:animation}, the topology updates preserve the vertex groups.

\begin{figure}
    \centering
    \includegraphics[width=\linewidth]{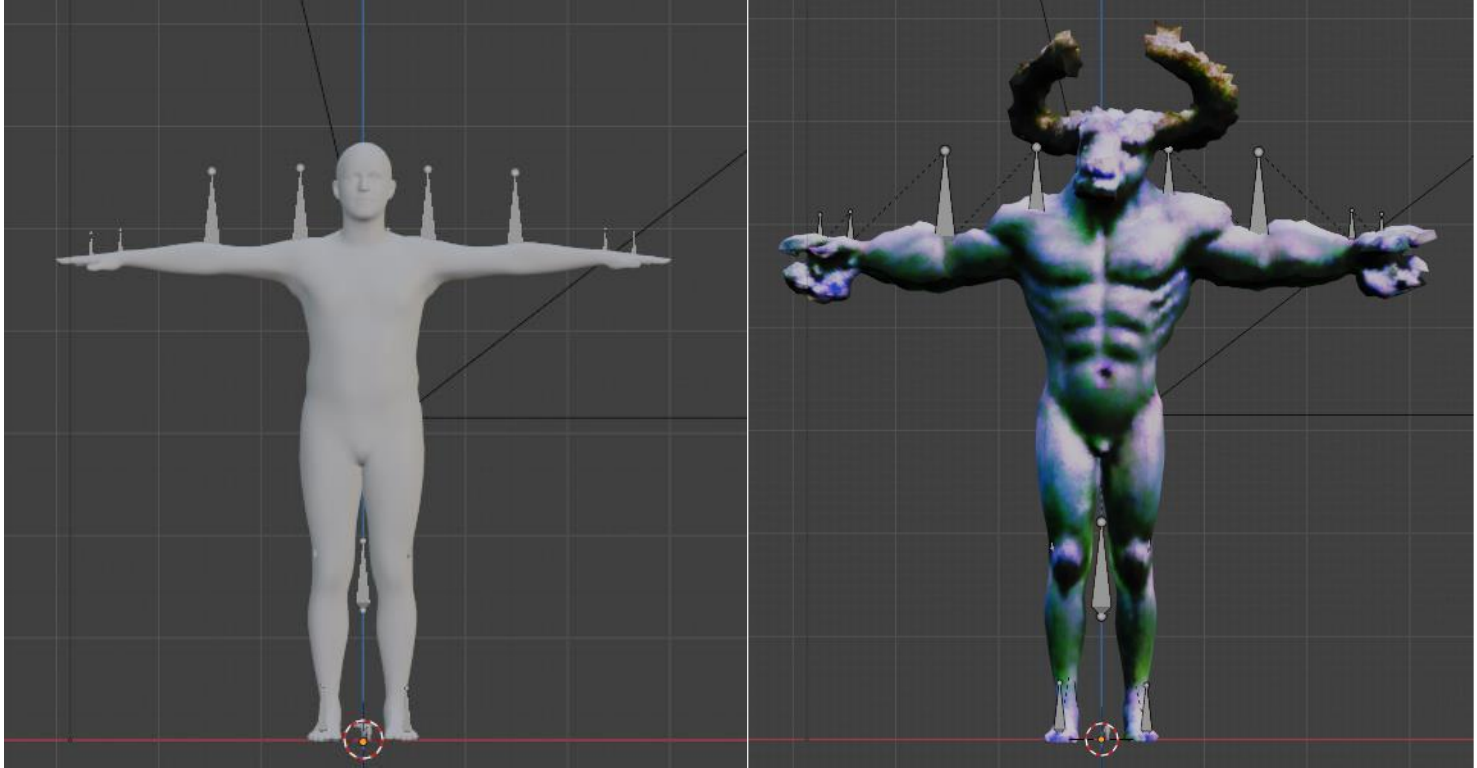}
    \caption{\bf{Animation Transfer.} \textmd{ We perform the animation transfer on the preexisting SMPL ~\cite{SMPL2015} animations, given in FBX format. the rig joints' position is given by the vertex groups, which are reasonably preserved due to the incremental nature of our mesh evolution process. We can then simply insert our mesh instead of the SMPL mesh and inherit existing animations for the given rig. Note, that in our experiments, for simplicity, we removed all existing blend shapes that are associated with the original vertex groups.}}
    \label{fig:animation2}
\end{figure}

\subsection{Application: Textured Meshes}
\begin{figure}
    \centering
    \includegraphics[width=\linewidth]{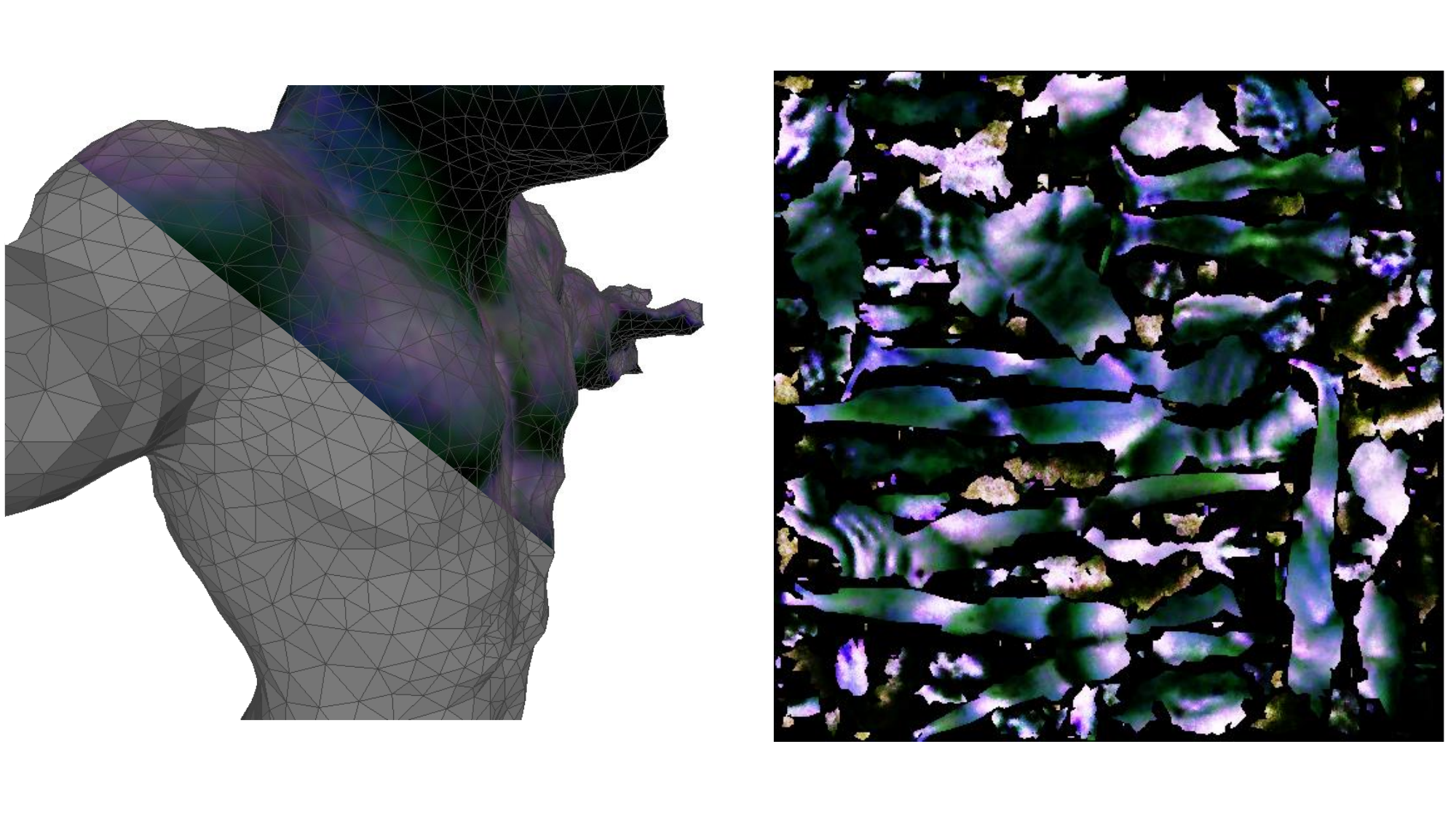}
    \includegraphics[width=\linewidth]{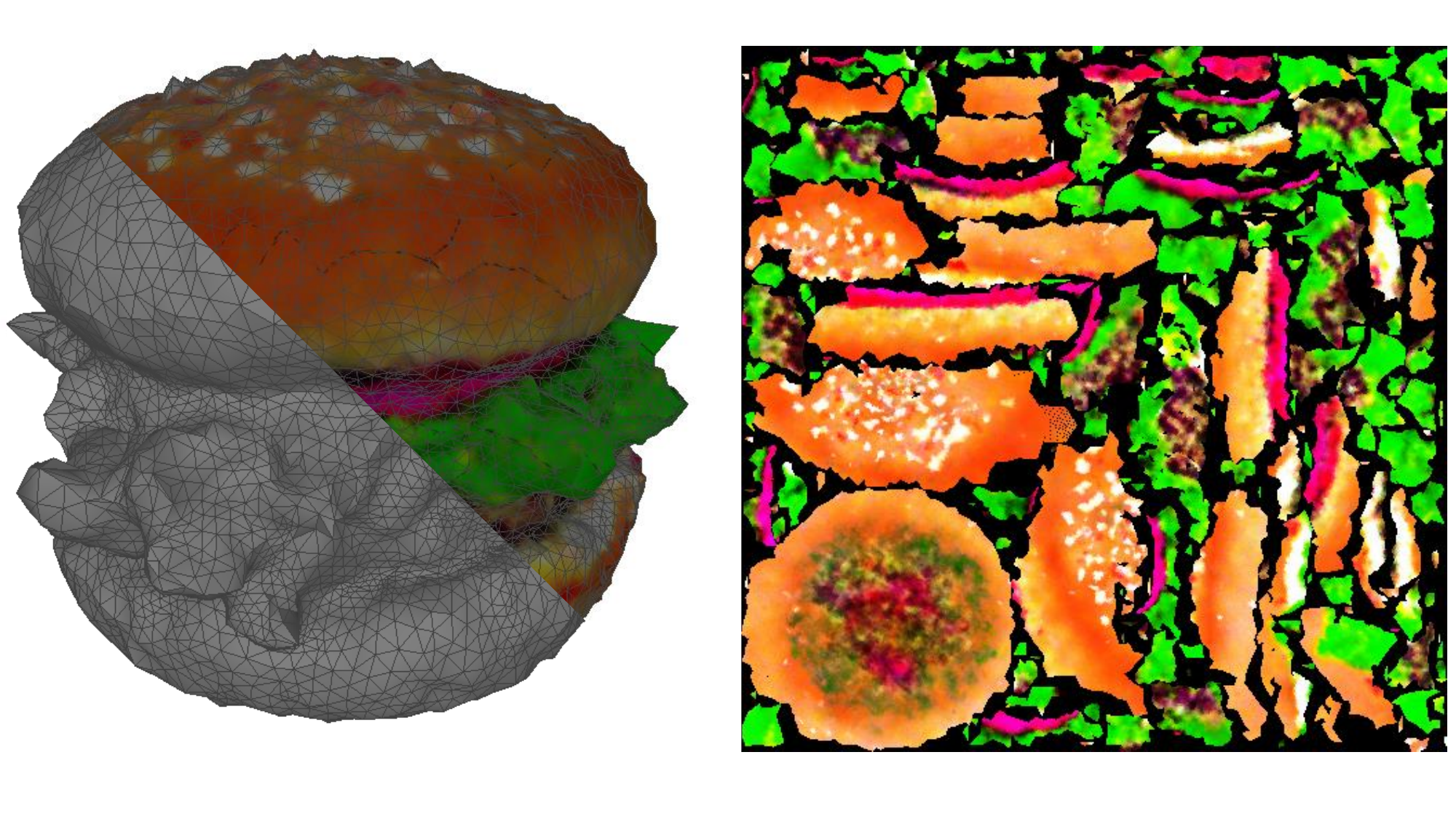}
    \caption{\textbf{Super-sampled colors to UV texture.} \textmd{One of the benefits of meshes is their ability to describe texture in a light and efficient-to-render way. The super-sampling operation allows to finely sample the color field over the mesh. This sampling can then be baked to a texture in conjunction with uv-unwrapping using Xatlas, to output a textured mesh to the user.}} 
    \label{fig:textured_meshes}
\end{figure}

As described in the main paper, in order to decouple the RGB rendering from the topology, our appearance network is queried on the super-sampled faces, in a way similar to that of Mesh colors \cite{meshcolors2010}.  UV textures are more common, and our super-sampled face colors can easily be turned into this representation, as shown in Figure~\ref{fig:textured_meshes}.

\bibliographystyle{ACM-Reference-Format}
\bibliography{sample-base}